\documentclass[twocolumn,prl,amssymb,amsmath,preprintnumbers,aps]{revtex4-1}
\usepackage{amssymb}
\usepackage{amsmath}
\usepackage{lineno}
\usepackage{bm,color,xcolor}
\usepackage{slashed} 
\usepackage{graphicx}
\usepackage{multirow}
\usepackage{comment}
\usepackage{mathtools}

\usepackage[colorlinks=true
,urlcolor=blue
,anchorcolor=blue
,citecolor=blue 
,filecolor=blue
,linkcolor=blue
,menucolor=blue
,linktocpage=true
,pdfproducer=medialab
,pdfa=true
]{hyperref}
\usepackage{lineno}

\newcommand{\beq}{\begin{equation}}
\newcommand{\eeq}{\end{equation}}
\newcommand{\bea}{\begin{eqnarray}}
\newcommand{\eea}{\end{eqnarray}}
\newcommand{\nn}{\nonumber\\}

\newcommand{\PL}{{\mathbb P}_L}

\newcommand{\st}{{\tilde t}}
\newcommand{\sT}{{\tilde T}}

\newcommand{\fbinv}{\ensuremath{\mathrm{fb}^{\mathrm{-1}}}}
\newcommand{\pt}{\ensuremath{p_{\mathrm{T}}}}
\newcommand{\vect}[1]{\boldsymbol{#1}}
\newcommand{\etmiss}{\ensuremath{E^{\text{miss}}_\mathrm{T}}}

\begin{document}

\title{\bf \Large Exploring Uncharted Soft Displaced Vertices in Open Data}

\author{ Haipeng An$^{\,(a,b,c)}$, Zhen Hu$^{\,(a,b)}$, Zhen Liu$^{\,(d)}$
, Daneng Yang$^{\,(a)}$}
 
\affiliation{(a)\it Department of Physics, Tsinghua University, Beijing 100084, China}

\affiliation{(b)\it Center for High Energy Physics, Tsinghua University, Beijing 100084, China}

\affiliation{(c)\it Center for High Energy Physics, Peking University, Beijing 100871, China}

\affiliation{(d)\it School of Physics and Astronomy, University of Minnesota, Minneapolis, MN 55455, USA}


\begin{abstract}
A cluster of {\it soft} displaced tracks corresponds to the dark matter co-annihilation regime. The long-lived regime is, in particular, motivated by the unexplored top partner physics. The background in this regime is extremely challenging to model using a traditional simulation method.
We demonstrate the feasibility of handling the formidable background using the CMS Open Data.
We perform this analysis to search for compressed and long-lived top partners in the 8 TeV CMS Open Data events with the integrated luminosity of 11.6 fb$^{-1}$ and obtain new limits. 
With 15-30 GeV mass splitting between the top partner and the DM candidate, we exclude the top partner mass below 350 GeV, which is more stringent than the ATLAS and CMS results using 8 TeV data with 20 fb$^{-1}$ luminosity. Our study also shows that the CMS Open Data can be a powerful tool to help physicists explore non-conventional new physics and even enable deriving new limits on exotic signals from data directly.
\end{abstract}

\maketitle

{\flushleft \bf Introduction---}
The presence of long-lived particles can provide striking signals in many new physics models~\cite{Barbier:2004ez,Giudice:1998bp,Meade:2010ji,Arvanitaki:2012ps,ArkaniHamed:2012gw,Grober:2014aha,Liu:2015bma,Chacko:2005pe,Burdman:2006tz,Kang:2008ea,Craig:2015pha,Davoli:2017swj,Liu:2019ayx,Hook:2019qoh}.
Requirement on the displaced decaying vertices associated with these particles offers an efficient and novel way to reject standard model (SM) background.
In many well-motivated cases, the long-lived particles decay into massive invisible particles, leaving only tracks of low energies~\cite{Grober:2014aha,Baker:2015qna}, which are hard to reconstruct.
%
This challenging signal can naturally arise from nearly degenerate two-level systems of a neutral Lightest Stable Particle (LSP) and a next-to-lightest stable particle (NLSP).
In particular, even when the splitting is as large as tens of GeV, the NLSP can still be long-lived,  
so long as the leading decay channel is through heavy SM states, e.g., top quark. In this case, 
the decay of the NLSP must go through a virtual top quark and a virtual $W$ boson. 
The LSPs in these models are usually SM singlets, which lead to inefficient annihilation in the early universe.
The nearly degenerate NLSP, on the other hand, opens up co-annihilation channels and restores the typical weakly interacting thermal dark matter paradigm~\cite{Griest:1990kh}.
Therefore, the presence of both particles offers a way to control and match the observed thermal dark matter abundance.

We consider two representative cases in this study. Case (A) is from supersymmetry (SUSY) with bino ($\chi^0_1$) as the LSP, and the stop (top squark $\tilde t_1$) as the NLSP, and the coupling of $\chi^0_1$ and $\tilde t_1$ to the top quark can be written as
\bea\label{eq:L}
{\cal L}_A = \left(- y_R \tilde t_1 \bar t_R \chi_1^0 - y_L \tilde t_1 \bar t_L \chi_1^0 \right) + {\rm h.c.} \ ,
\eea
where $y_R = - (2\sqrt{2}/3) g_1 \sin\theta_{t}$ and $y_L = (\sqrt{2}/6) g_1 \cos\theta_{t}$ with $g_1$ the hypercharge coupling and the $\theta_t$ the mixing angle of the two top partners. In case (B), the model we consider is a simplified model with the LSP a vector dark matter state and the NLSP a colored Dirac spinor. This model can be traced to extra dimension models with KK-parity or the little Higgs models with T-parity. Therefore, we call the LSP $A_H$ and the NLSP $t_H$. The couplings of $A_H$ and $t_H$ to the top quark can be written as
\bea\label{eq:LB}
{\cal L}_B = \left(- g_R \bar t_H \gamma_\mu t_R - g_L \bar t_H \gamma_\mu t_L \right) A_H^\mu + {\rm h.c.}\ ,
\eea
where $g_R = g_H\sin\theta_t$, $g_L = g_H\cos\theta_t$.
The typical value of $g_H$ is around $g_1$.
In both cases (A) and (B), to avoid overproducing the dark matter in the early universe, $\Delta$ is constrained to less than about 40 GeV~\cite{deSimone:2014pda,Ellis:2014ipa,Ellis:2014ipa,Ibarra:2015nca,Liew:2016hqo,Mitridate:2017izz,Keung:2017kot,Pierce:2017suq,Keung:2017kot,Ellis:2018jyl,Garny:2018icg,Biondini:2018ovz}.

In SUSY models, the top squark mass should be less than about 1 TeV to effectively solve the fine-tuning problem~\cite{Papucci:2011wy}.
However, when $m_{\tilde t_1} > m_{\chi^0_1} + m_t$,
traditional searches for the top squark, which consider signals composed of energetic tracks of hadronic or leptonic type, plus missing transverse energy (MET), have excluded the bulk parameter region for this possibility~\cite{Aad:2020sgw,ATLAS:2020llc,ATLAS:2020dav,Sirunyan:2019glc,Sirunyan:2019xwh,Sirunyan:2019ctn}, making the complementary parameter region more interesting. There are fewer studies for the extra dimension models or the little Higgs models. However, with the KK-parity or the T-parity, the signal is similar, and if we recast the SUSY result to these models, the constraints on the mass of the top partner in the bulk region are also TeV scale. As a result, in these theories, the compressed region, where the LSP and NLSP have small mass gaps, is motivated to produce the electroweak scale naturally with the current LHC data.

\begin{figure}[tb]
\centering
\includegraphics[width=8.2cm]{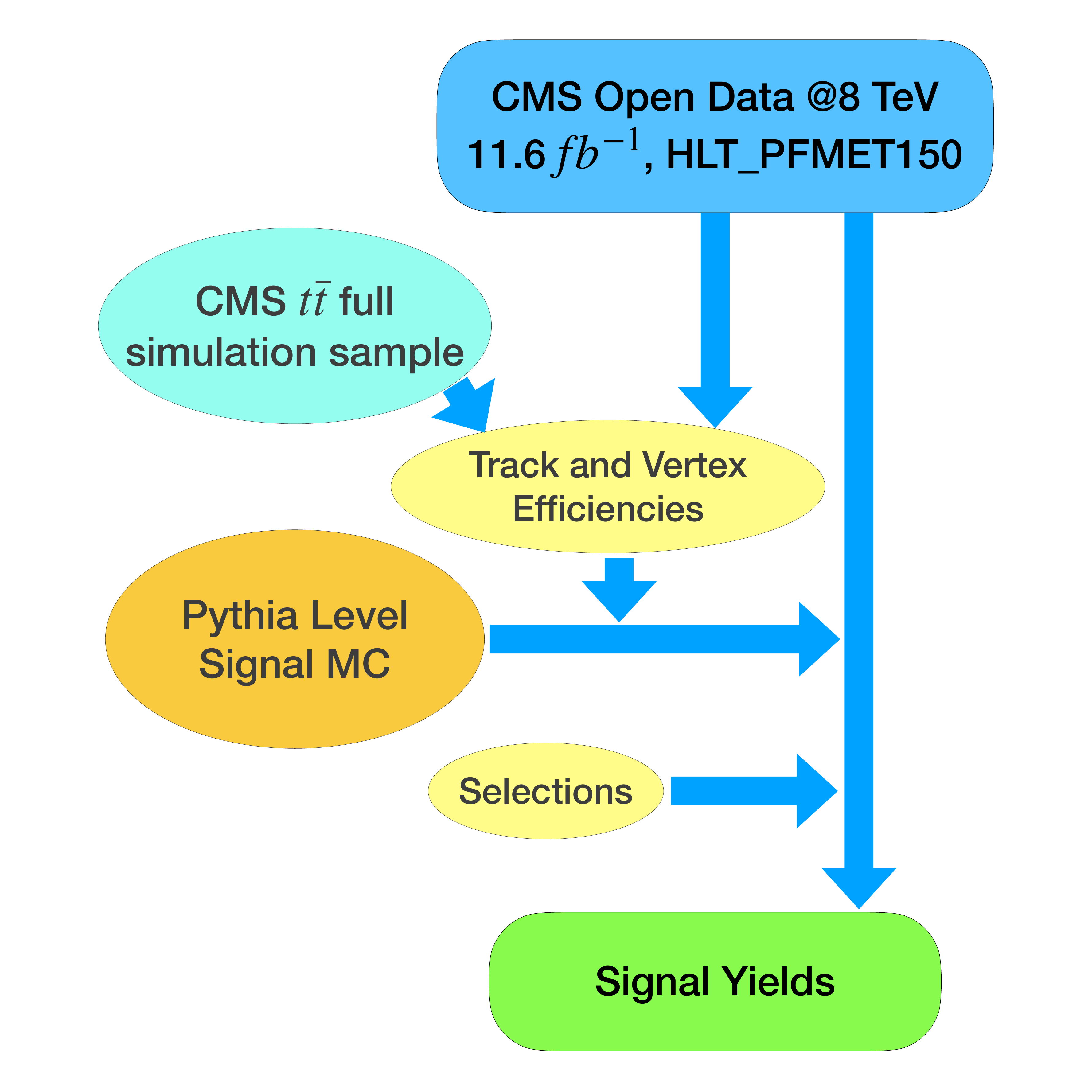}
\caption{\label{fig:chart}
Schematic flowchart of our analysis.
}
\end{figure}

In this letter, we propose to search for the degenerate region of the two-level system by looking for signals with displaced soft tracks and large MET signatures. The complex experimental environment and appearance of nonstandard background prevents a direct simulation-based search for this challenging signal. We hence take a data-driven method. 
We design a hybrid analysis and validation chain between experimental data, full simulation data, and fast simulation data to carry out such an analysis. 
A schematic presentation of our analysis method is shown in Fig.~\ref{fig:chart}.
We perform a vertex reconstruction based on all the selected displaced tracks and search for signal-like vertices from the products.
By selecting displaced tracks and vertices, one can suppress prompt decaying backgrounds. 
The backgrounds associated with intermediate B mesons can still produce displaced vertices.
However, the QCD-induced ones are usually soft, being suppressed by the requirement of monojet and MET.
The B mesons from top decays are inherently energetic and vetoed by selecting only soft tracks using a jet level upper-cut on the $H_T^V$, which is defined as the scalar sum of the $\pt$s of the tracks associated with the reconstructed vertex.

With our analysis, we set limits on the signal in a conservative statistical approach, based on the 11.6 fb$^{-1}$ collision data collected by the CMS experiment at 8 TeV.
Compared to the other 8 TeV results, we demonstrate that our method is significantly advantageous in the compressed region with $15 < \Delta < 30$ GeV.
The idea of our proposed new search can be further extended to a general class of long-lived particle searches, 
exploring the soft limit for the displaced vertex signatures, covering the blind spots of current searches
. We hope our study demonstrates and accelerates the new era of open science for particle physics~\cite{Larkoski:2017bvj,Tripathee:2017ybi,Chen:2018drk,Andrews:2019faz,Komiske:2019jim,Cesarotti:2019nax,Apyan:2019ybx,Mccauley:2019uis,Abdallah:2020pec,Lassila-Perini:2021xzn}.

\begin{figure}[tb]
\centering
\includegraphics[width=3.2in]{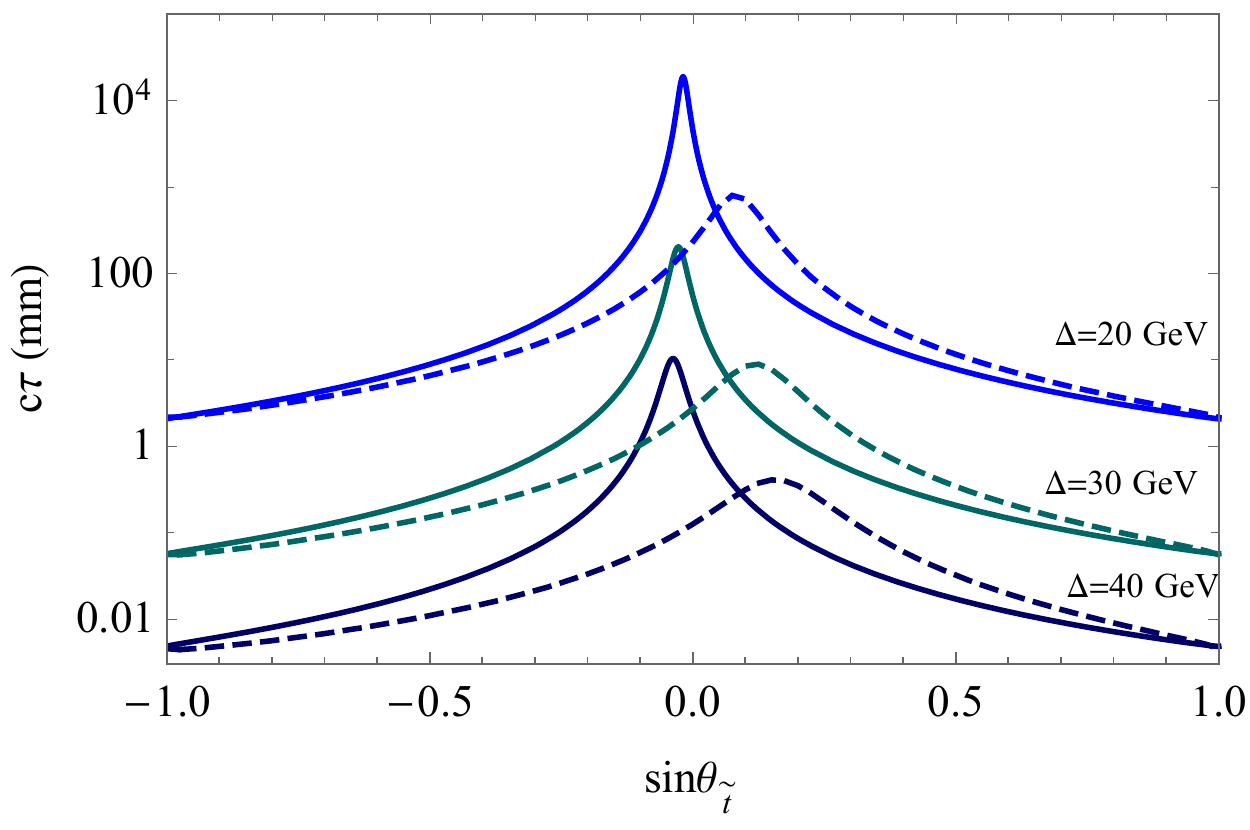}
\caption{
Proper lifetime $c\tau$ of top partners as functions of $\sin\theta_{t}$ for $500$ GeV top partner mass. The blue, green, and black curves correspond to $\Delta = 20, 30$ and 40~GeV. The solid curves show results for the SUSY model (A), with one-loop QCD RG corrections to the righthanded operators included. The dashed curves are the tree-level results for model (B).
}
\label{fig:ctau}
\end{figure}

{\flushleft \bf NLSP decays---}
From the Lagrangians for cases (A) and (B), one can derive the tree-level effective Lagrangians that governs the decay of the NLPs.
\bea
{\cal L}_{A,B}^{\rm tr} = C^R_{A,B} {\cal O}^R_{A,B} + C^L_{A,B} {\cal O}^L_{A,B} \ ,
\eea
where for the SUSY case (A) the effective operators are
\bea
{\cal O}^R_A &=& (\bar f' \gamma_\mu \PL f) (\bar b \gamma^\mu \PL \chi \tilde t_1) \ , \nn
{\cal O}^L_A &=& (\bar f' \gamma_\mu \PL f) (\bar b \gamma^\mu \PL i \!\not\!\!D (\chi \tilde t_1)) \ .
\eea
The corresponding Wilson coefficients are
\bea
C^R_A = - \frac{ y_R g_2^2}{2 m_W^2 m_t} \ , \;\; C^L_A = \frac{y_L g_2^2}{2 m_W^2 m_t^2} \ .
\eea
For case (B) we have
\bea
{\cal O}^R_B &=& (\bar f' \gamma_\mu \PL f) (\bar b \gamma^\mu \PL \!\not\!\! A_H t_H) \ , \nn
{\cal O}^L_B &=& (\bar f' \gamma_\mu \PL f) (\bar b \gamma^\mu \PL \!\not\!\! D (\!\not\!\! A_H t_H) ) \ ,
\eea
with the Wilson coefficients
\bea
C^R_B = - \frac{g_R g_2^2}{2 m_W^2 m_t} \ , \;\; C^L_B = \frac{g_L g_2^2}{2 m_W^2 m_t^2} \ .
\eea
In both cases, the right-handed operators ${\cal O}^R_{A,B}$ are dimension seven, whereas the left-handed operators are dimension eight. Therefore, compared to the right handed operators, the contributions from left handed operators are suppressed by $\Delta/m_t$, which is about $1/6$ in the parameter region for this study. Especially in case (A), $y_R/y_L = -4 \tan\theta_{\tilde t}$. Therefore as long as $\tan\theta_{\tilde t} > {\cal O}(0.1)$ the decay of $\tilde t_1$ is mostly contributed by the right-handed operators. Indeed, as we can see from Fig.~\ref{fig:ctau} that if $\tilde t_1$ is purely $\st_L$, its lifetime is approximately ${\cal O}(10^3)$ times larger than the pure $\st_R$ case.
On the other hand, if the NLSP is mostly left-handed top partners, one usually expects a bottom partner to contribute to the optimal channel for collider search. The $c\tau$ values for case (B) are shown in Fig.~\ref{fig:ctau}. One can see that for typical choices of $g_{R,L}$ the behavior of the $c\tau$ in case (B) is almost identical to case (A). Thus, the same search strategy applies to both cases. Therefore, in the following discussions, we focus on the SUSY case (A) to illustrate our search strategy.

When $\tilde t_1$ decays, the typical energy scale in the internal virtual particles is around $\Delta$.
However, to get the effective operators, we encounter energy scales of $m_{\tilde t_1}$, $m_t$, and $m_W$, which are much larger than the minor splitting $\Delta$.
Therefore, we include the leading log corrections to the right-handed operators to calculate the decay rate and the leptonic and hadronic branching ratios.
To calculate the leading part of the QCD corrections, we use the renormalization group method to crank down the energy scale from $m_{\tilde t}$ to $\Delta$.
We use the non-relativistic quantum field theory technique, similar to the heavy quark effective theory~\cite{Manohar:2000dt}. The detailed one-loop calculations are presented in the supplemental material.

{\flushleft \bf Data processing---}
We apply a single un-prescaled trigger HLT\_PFMET150 for the 2012 MET primary dataset~\cite{CMSMET12}, which includes valid runs from CMS RunB, RunC and corresponds to an integrated luminosity of $11.6$ \fbinv~\cite{CMSdata12}. 
The data contains high-level physics objects and additional event- and sector-based information that allows for offline corrections and kinematic refitting.
The analysis is performed using the CMS SoftWare ({\sc CMSSW}) version 5.3.32, provided by the Docker image released through the CMS Open Data portal with build-in tools for offline analysis~\cite{CMSWB}.

We use reconstructed candidates in the Analysis Object
Data (AOD). 
Primary Vertices (PV) correspond to those lying on the beamline, including those from pile-up events. 
Jets, together with the missing transverse momentum vector, are reconstructed using the particle flow (PF) algorithm~\cite{CMS-PAS-PFT-09-001,CMS-PAS-PFT-10-001,Sirunyan_2017}.
Besides jets and MET, our analysis is mainly based on collections of reconstructed tracks.
We use only the high purity tracks with track $\pt>1$~GeV and $|\eta|<2.4$ to ensure a high quality~\cite{Collaboration_2014}.
For each track, the transverse impact parameter $d_{xy}$ is computed from the beam-spot, which has been corrected in each data-taking period. 
The track four-momentum is computed at the vertex position, assuming the mass of a charged pion.

Displaced vertices are not provided in the AOD sample, and we reconstruct them offline. 
We use the build-in Trimmed Kalman vertex filter and finder from the {\sc CMSSW} to fit and find vertices from displaced tracks~\cite{Collaboration_2007}. 
We select displaced tracks with $|d_{xy}|/\sigma_{d_{xy}}\geq 4$
and valid vertices with $\chi^2>0$, $ndf>0$, $\chi^2/ndf < 5$. Here $d_{xy}$ is the fitted vertex displacement in the transverse plane,  $\sigma_{d_{xy}}$ the uncertainty of the fit, and $ndf$ the number of degrees of freedom.
To avoid contamination from interactions with the beam pipe or detector material, we further require the reconstructed vertices to have a distance to the beamline $d_{BV}$ from $0.1$ mm to $20$ mm~\cite{Sirunyan:2018pwn}. 
This {\it very conservative} upper cut in transverse displacement is imposed mainly to avoid extrapolation of detector and background estimation from data, given the lack of data and full-simulation data in this regime. This conservative choice also simplifies our study in that the signal tracks can be regarded as coming from a single high purity category~\cite{Collaboration_2014}. 
%

{\flushleft \bf Simulations and efficiencies---} The signal pairly-produced top squark events are generated using {\sc MadGraph5}~\cite{Alwall:2014hca} with up to two associated jets, with our benchmark model parameters using the build-in MSSM model file~\cite{deAquino:2011ub}.
The generated parton-level events are interfaced to {\sc Pythia8}~\cite{Sjostrand:2007gs,Sj_strand_2006} for parton shower and R-hadron decays, where an MLM matching is performed 
and the jets are reconstructed using the {\sc FASTJET3} package~\cite{Cacciari:2011ma,Cacciari:2005hq}. The Pythia-level events are reweighted with matrix element square to include kinetic information of the $\st_1$ decay. 
We identify the intermediate top squarks, B mesons and randomly displace the vertex positions for each simulated event, following the decay local probability distribution based on the event-by-event lab-frame lifetime and velocities.


The signal yield under event reconstruction and selection is computed based on a mixture of requirements on the simulated kinematic variables, and on applying dedicated track and vertex efficiencies. 
The track efficiency is used to determine the number of displaced tracks under the selection $|d_{xy}|/\sigma_{d_{xy}}\geq 4$ that are available to vertex reconstruction. 
The vertex efficiency determines if a signal vertex of a given number of displaced tracks can be reconstructed by the detector. 
We expect the inclusion of the two efficiencies to capture most of the detector effect, making our signal comparable to the data. 

To obtain the $\sigma_{d_{xy}}$ distribution of the tracks from the stop and B-meson decay,  
we consider a set of high fidelity tracks selected with $|d_{xy}|<250~\rm\mu m$ and $|\eta|<2.4$ in the Open Data. 
For a simulated track with $\pt^0$ and $d_{xy}^0$, we select the Open Data tracks in the corresponding $p_T$ bin and compute the efficiency as the fraction of the number of tracks satisfying $\sigma_{d_{xy}}<|d_{xy}^0|/4$. 
Besides the selection efficiency, we also include a constant 90\% reconstruction efficiency for each track according to the CMS track performance~\cite{Collaboration_2014}. 

To estimate the displaced vertex identification and reconstruction efficiency (hereafter vertex efficiency), we exploit the MC truth information from an 8 TeV full-detector simulated $t\bar{t}$ sample obtained from the CMS Open Data portal~\cite{CMSTTJets12}.
In this $t\bar{t}$ sample, we select hadronically decaying $B_0$ or $\bar{B}_0$ mesons by requiring the energies to be within 10 to 30 GeV, and the vertex position to be larger than $0.5~$mm but smaller than $18~$mm from the beam line.
These $B_0/\bar{B}_0$ meson candidates have properties mimicking our signals; hence can be used to calculate the vertex efficiency. 
We further categorize these reconstructed displaced candidates by the number of displaced tracks $N_{gen,tk}$, with each track passing $d_{xy}>0.5$ mm, $\pt>1$ GeV, $|\eta|<2.4$ and are not coming from tertiary vertices. 
In each catalog, the vertex efficiency can be computed by matching the reconstructed displaced vertices or their associated tracks to the simulation truth-level ones. 
We use two methods to perform this displaced vertices/tracks matching. 
The first method is based on the fraction of tracks in a reconstructed vertex that can be matched to the truth-level ones originating from the decaying vertex.
The second method is based on a requirement of the separation between the reconstructed and the truth-level vertex. 
We found the matched candidates from the two methods have $\gtrsim 60$\% overlap in all considered catalogs.
We use the results of the first one for other parts of the analysis. We describe details of the efficiency extraction in the supplemental material.

\begin{figure}[tb]
  \centering
  \includegraphics[width=9.5cm]{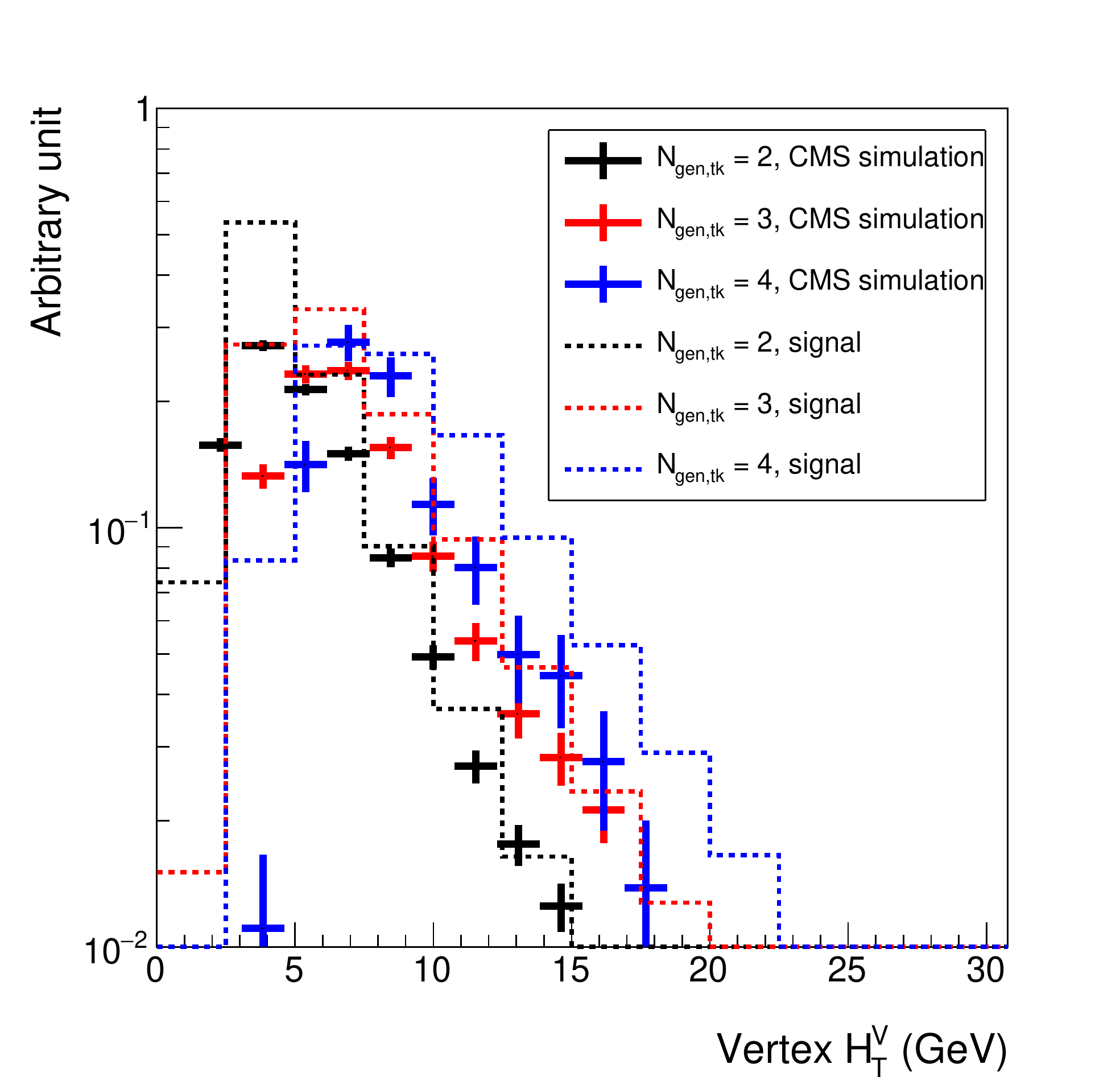}
\caption{\label{fig:vtxff} 
Vertex $H_T^V$ distributions for templates of the CMS simulation (bold bars) and of our signal sample with $\Delta=20~$GeV, $m_{\tilde{t}_1}=360~$GeV.
The agreement between the two simulations verifies our modeling of signal efficiencies under selection and reconstruction. 
}
\end{figure}

In our signal events, because the B-meson comes from long-lived R-hadron decays, there are tertiary structures in the displaced vertices, depending on the flying distance of the B-meson. 
To properly account for this effect, we merge nearby {\sc pythia}-level vertices that are below the reconstruction resolution before applying the vertex efficiency. 
Details of the vertex combination are contained in the supplemental material. 
When applying the efficiencies, we note that vertices with $N_{{\sc pythia},tk}$ equals two or three have a probability of passing our data selection $N_{vtx,tk}\geq$4. 
This is caused by contaminating tracks from other vertices.

After merging nearby vertices, we compare the vertex $H_T^V$ distribution of the signal-like templates from the CMS simulation and a benchmark signal sample with $\Delta=20~$GeV and $m_{\tilde{t}_1}=360~$GeV. This comparison is performed in a one-vertex control region, defined by requiring one displaced vertex with $N_{vtx,tk}\geq$4, MET larger than 150 GeV, and transverse momentum of the ISR jet higher than 150 GeV.
As shown in Fig.~\ref{fig:vtxff}, the selected templates do resemble the simulated signal samples in the $N_{gen,tk}=2,3,4$ categories. 

\begin{figure*}[tb]
  \centering
  \includegraphics[width=8cm]{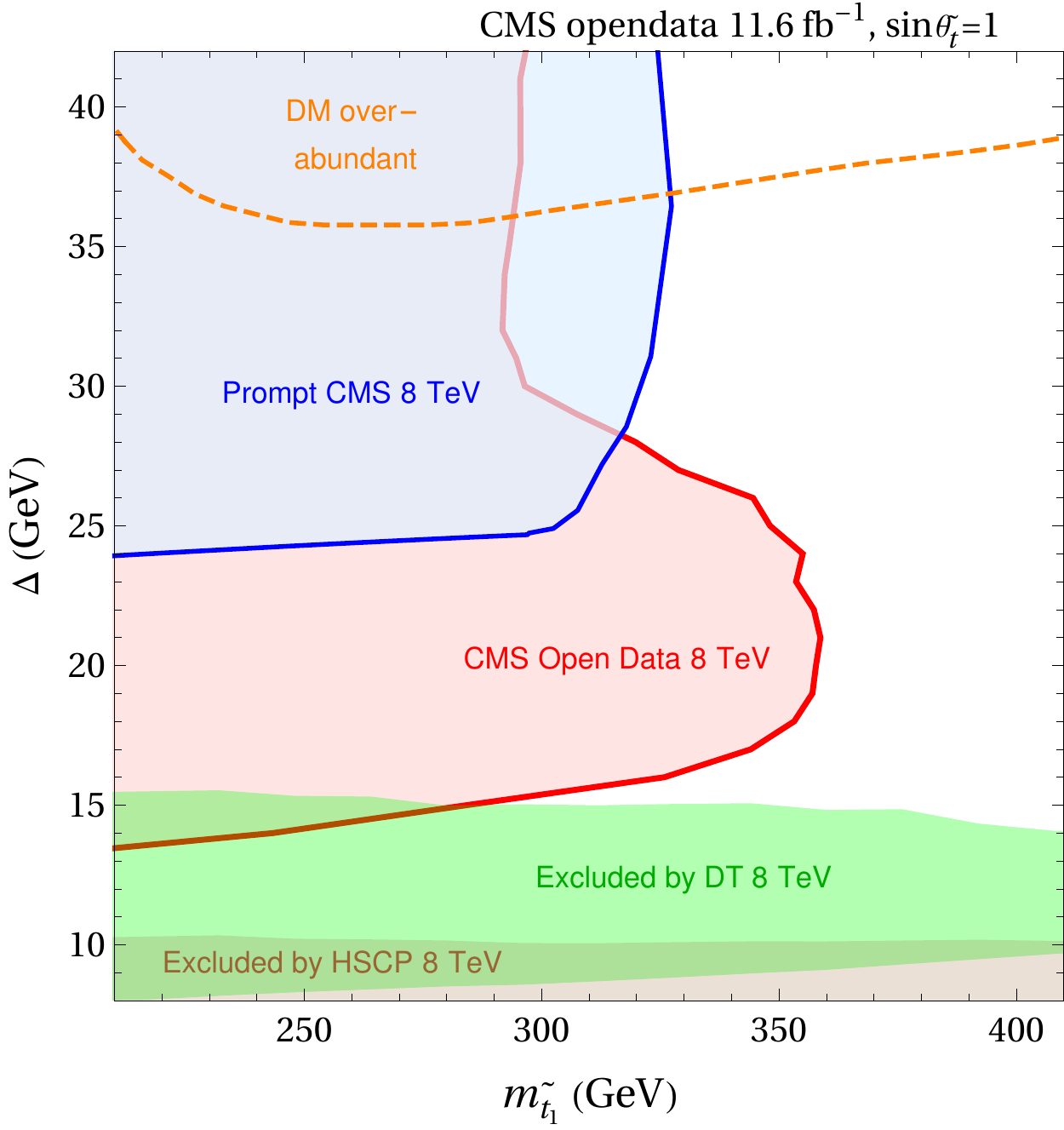}
  \includegraphics[width=8cm]{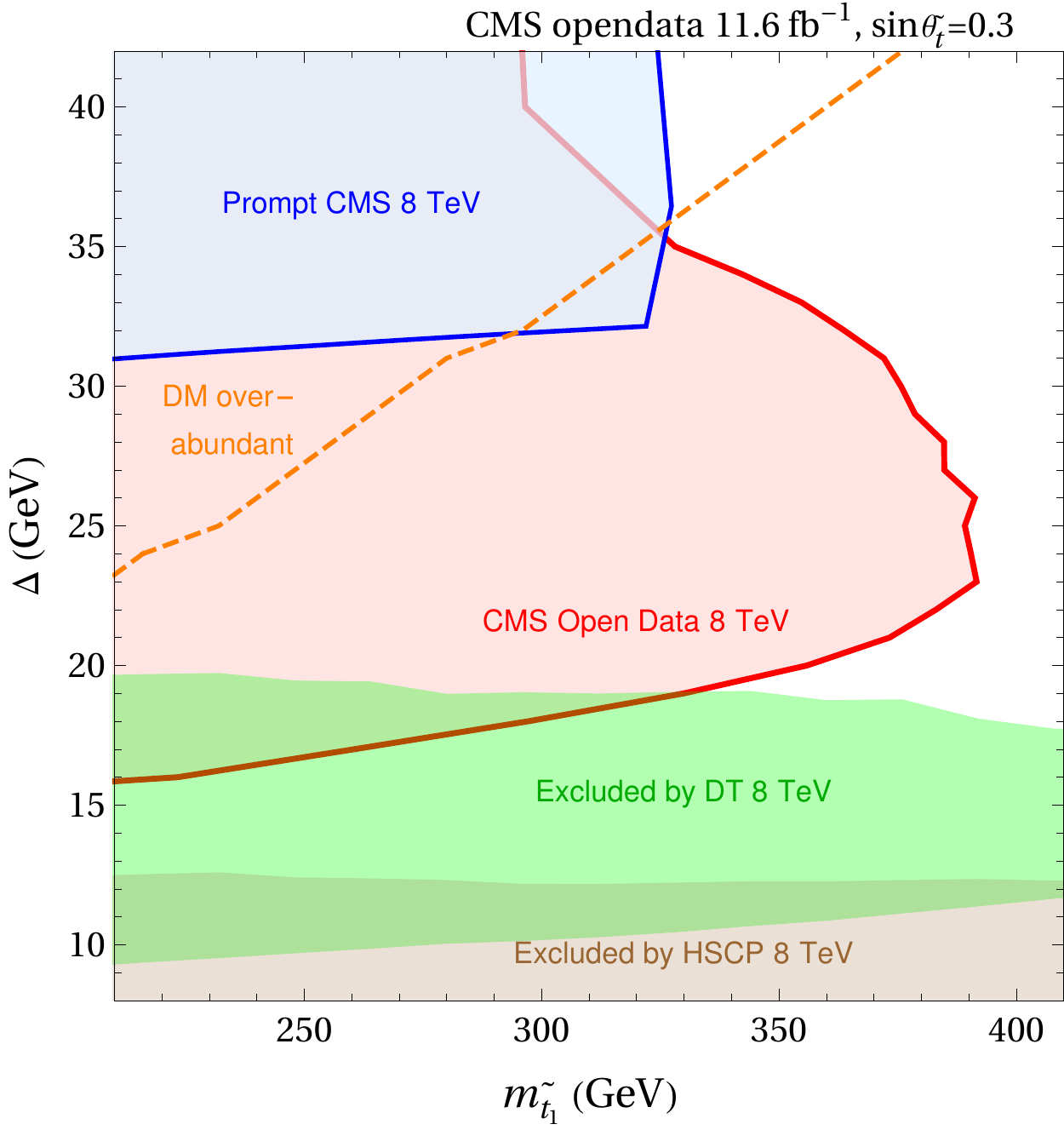}
\caption{\label{fig:limits} Limits on the top squark mass and the mass splitting. 
CMS 8 TeV limits from Ref.~\cite{Khachatryan:2015pot}($\tilde{t}\to b l \bar{l} \tilde{\chi}_1^0$ channel). On the left panel, the DM relic density is taken from Ref.~\cite{Keung:2017kot} with the bound state effects included. 
On the right panel, the contour for the DM relic density is computed using the {\sc MadDM} program~\cite{Backovic:2013dpa,Arina:2020kko}, considering $\sin\theta_{\tilde{t}}=0.3$ and sbottom mass 1 TeV. 
The 8 TeV disappearing track (DT) searches~\cite{Aad:2013yna,CMS:2014gxa} and heavy stable charged particle (HSCP) searches~\cite{ATLAS:2014fka,Chatrchyan:2013oca} are re-interpreted from the existing searches~\cite{Chatrchyan:2013oca} using the framework in Ref.~\cite{Liu:2015bma}. 
For a fair comparison, given the limited 8 TeV open data energy and luminosity, we only include the current search results with 8 TeV data. 
For a comparative understanding regarding the 13 TeV searches, please see the discussion in the text and the supplemental material.
}
\end{figure*}

{\flushleft \bf Results---} We select events with $\pt^{j1} >300$ GeV, $\etmiss>300$ GeV, and two displaced vertices with $N_{vtx,tk}\geq$4 and $H^V_{\rm T}<40$~GeV. A selection flow is shown in Table~\ref{tab:cutflow} for the data and a signal benchmark with $m_{\tilde{t}_1}=360~$GeV and $\Delta=20~$GeV. In the table, we include the intermediate steps to demonstrate the efficiency loss under the requirement of $N_{vtx,tk}$. 
With higher statistics, one may require tighter $H^V_{\rm T}$ for improvement.
Based on the selected events, we set limits on the signal strength using the Feldman-Cousins method~\cite{PhysRevD.57.3873}.
Our results are shown in Fig.~\ref{fig:limits}. We find a 68\% (95\%) confidence level (CL) limit corresponds to 1.28 (3.09) signal events with the zero background post-selection.


\begin{table}[tb]
\begin{center}
\begin{tabular}{c|c|c}
\hline
\hline
Selection                                         & Data        & Signal BM \\
\hline
MET primary                                       & 4.3$\times 10^7$   &  -    \\
$\pt^{j1} >300$ GeV, $\etmiss>300$ GeV            & 2.3$\times 10^5$   &  198  \\ 
One displaced vertex ($N_{vtx,tk}\geq$2)          & 4.7$\times 10^4$   &  83  \\ 
One displaced vertex ($N_{vtx,tk}\geq$3)          & 3.8$\times 10^3$   &  65    \\ 
One displaced vertex ($N_{vtx,tk}\geq$4)          & 3.0$\times 10^2$   &  39  \\ 
\hline
One displaced vertex with $H^V_{\rm T}<40$~GeV    &    245             &  39   \\ 
Two displaced vertices with $H^V_{\rm T}<40$~GeV  &          0         &  3.0  \\ 
\hline
\hline
\end{tabular}
\caption{\label{tab:cutflow} Selection flow table. Each selection criterion is applied on top of the previous ones. A horizontal line is added to separate the selections with/without the $H^V_{\rm T}$ requirement. The MC benchmark corresponds to a top squark mass of 360 GeV and a mass gap of 20 GeV.
}
\end{center}
\end{table}



In the compressed region, the traditional search based upon kinematic reconstructions, such as MT2 variables, in the top squark pair production channels become insensitive, as most of the energy of the top squarks is carried away by the neutralinos and observable tracks are very soft. 
Nevertheless, relying on the initial state radiation (ISR), both CMS and ATLAS conducted searches for missing energy plus jets~\cite{Sirunyan:2018omt,Sirunyan:2017wif,Aad:2020sgw}. 
However, in these studies, the decay of $\st_1$ is assumed to be prompt. Therefore, these results do not fully apply to the $\st_1$-bino co-annihilation model. We show in blue-shaded regions in Fig.~\ref{fig:limits} constraints from these prompt searches, where we cut away their coverage in parameter regions where the top squark proper lifetime is beyond 0.2~mm. Note that the lifetime changes as eighth power of the mass splitting, hence the prompt searches lose sensitivity rapidly as the splitting decreases. Given the exact reconstruction for the prompt object in the search is not publicly available, we choose 0.2~mm as the cut for prompt searches.\footnote{Note that the prompt search at requires the leptons have an impact parameter less than 0.1~mm, which is well-contained by our lifetime smaller than 0.2~mm cut. In addition, the search uses a vetoing of b-tagged jets, which is hard to model without knowing the full detail of the tagging algorithm, especially for further displaced b-jets.}
In the regime with larger splittings, e.g., mass splitting more than 30 GeV, our search sensitivity is driven by the displacements from B-meson with prompt top squark decays, approaching that of the CMS 8 TeV analysis. Note that our limit in the prompt region is weaker than that of the CMS 8 TeV analysis.
In the compressed regime with $\Delta<30$, our method significantly improves the sensitivity and excludes $m_{\tilde{t}}$ up to 350 (380) GeV for $\sin\theta_t=1$ ($\sin\theta_t=0.3$). 

Various long-lived particle search channels at the LHC cover complementary regions. Based upon the framework established in Ref.~\cite{Liu:2015bma}, we implement and recast the relevant search sensitivities for compressed stops. In the parameter regions of benchmark models considered in this paper, the heavy stable charged particle searches and disappearing track searches by both ATLAS~\cite{ATLAS:2014fka,Aad:2013yna} and CMS~\cite{Chatrchyan:2013oca,CMS:2014gxa} are relevant and show complementary coverage. We show in Fig.~\ref{fig:limits} the reinterpreted limits in our model from these 8 TeV long-lived particle searches. We can see that the prompt and long-lived searches leave a critical gap in the stop-bino mass plane, in particular, motivated by the relic abundance considerations. The displaced vertex dijet search at 8 TeV~\cite{CMS:2014wda}, which requires a pair of 30 GeV $p_T$ jets,  is inefficient due to the soft nature of our signal. Hence we did not recast it here. Our proposed search, validated by the CMS Open Data, already shows stronger constraints in the motivated regions, highlighting the importance of searches for soft hadronic long-lived signatures.

{\noindent \bf Summary and outlook---} 
In this work, we point out a well-motivated and yet challenging regime of collider searches for new physics, namely, long-lived {\it soft} and {\it hadronic} signatures. This region is particularly motivated by a broad class of dark matter co-annihilation considerations. Yet, the softness and hadronic nature make it a very challenging signal to look for at the LHC. Simultaneously, due to the insufficient understanding and modeling of the subtle background for them, it is preposterous for theorists to estimate how viable the search would be at the LHC. Despite the difficulties, we directly confront the challenge, beyond motivating the signatures, using the CMS Open Data and simulation framework to understand the signal and background behaviors. 

To provide a concrete BSM model example, we consider the long-lived nature for a SUSY-like signal and a composite-model-like signal from top partners as the NLSP. As one of the key predictions in generically natural models, the top partners can be long-lived in the compressed regime. Moreover, this regime also provides a viable LSP DM candidate via the co-annihilation mechanism.
We are motivated to fully explore such possibilities by proposing a new long-lived particle search that is poorly covered by conventional mono-jet searches and long-lived particle searches. 

We propose to search for displaced vertices within the MET-triggered samples. 
We showed its excellent coverage in parameter space, complimenting current prompt searches and long-lived signature searches. 
Our proposed search achieves better coverage in the long-live regime and enjoys a great prospect to improve with higher and higher luminosities from its low-background and statistical-limited nature. This is in contrast to mono-jet search dominated by systematic uncertainties. 
The idea of our proposed new search can be further extended to a general class of long-lived particle searches, 
exploring the soft limit for the displaced vertex signatures, covering the blind spots of current searches
. We hope our study demonstrates and accelerates the new era of open science for particle physics.

{\noindent\bf Acknowledgment ---}
We thank CERN, the CMS collaboration, and the CMS Data Preservation and Open Access (DPOA) team for making the LHC data accessible. We thank Alexander Ridgway for the work at the beginning stage of this project. 
We thank Jesse Thaler for helpful comments and suggestions. 
HA is supported by NSFC under Grant No. 11975134, the National Key Research and Development Program of China under Grant No.2017YFA0402204 and the Tsinghua University Initiative Scientific Research Program. ZH is supported by NSFC under Grant No. 11975010 and No. 12061141002, and the Tsinghua University Initiative Scientific Research Program. ZL acknowledges the Minnesota Supercomputing Institute (MSI) at the University of Minnesota for providing resources that contributed to the research results reported within this paper.


\bibliographystyle{utphys}
\bibliography{reference}


\onecolumngrid
\appendix



\newpage

\begin{center}
{\bf \Large Supplemental Materials}
\end{center}

\section{Quantum corrections to the $\st_1$ decay process}
\label{sec:loop}

The effective Lagrangian determines the $\st_1$ decay can be written as
\bea
{\cal L}^{\rm tr} = C_R^{(\ell)} {\cal O}_R^{(\ell)} + C_R^{(h)} {\cal O}_R^{(h)} +  C_L^{(\ell)} {\cal O}_L^{(\ell)} + C_L^{(h)} {\cal O}_L^{(h)}\ ,
\eea 
where the superscript $\ell$ $(h)$ are for leptonic (hadronic) decay channels. From perturbative calculation it is easy to show that 
\bea
&&{\cal O}_R^{(h)} = (\bar u_i \gamma_\mu \PL d_j) V_{ij} (\bar b \gamma^\mu \PL \chi \tilde t_1),\\
&&{\cal O}_R^{(\ell)} = (\bar \nu_i \gamma_\mu \PL l_j)  (\bar b \gamma^\mu \PL \chi \tilde t_1) \,\\
&&{\cal O}_L^{(h)} = (\bar u_i \gamma_\mu \PL d_j) V_{ij} (\bar b \gamma^\mu \PL i \!\not\!\!D (\chi \tilde t_1)) \,\\
&&{\cal O}_L^{(\ell)} = (\bar \nu_i \gamma_\mu \PL l_j)  (\bar b \gamma^\mu \PL i \!\not\!\!D (\chi \tilde t_1)) \ ,
\eea 
where $V_{ij}$ is the CKM matrix, and $D$ is the covariant derivative for the strong interaction.\footnote{The summation of color indices is also implicit. The color structure of operators become important for loop corrections that we include in this study.} At tree level the Wilson coefficients are
\bea
C_R^{(h)} = C_R^{(\ell)} = - \frac{ y_R g_2^2}{2 m_W^2 m_t} \ , \;\; C_L^{(h)} = C_L^{(\ell)} = \frac{y_L g_2^2}{2 m_W^2 m_t^2} \ .
\eea
The right-handed operators ${\cal O}_R^{(h)}$ and ${\cal O}_R^{(\ell)}$ are dimension seven whereas the left-handed operators ${\cal O}_L^{(h)}$ and ${\cal O}_L^{(\ell)}$ are dimension eight. 

Neglecting the masses of the light quarks and leptons, as well as the bottom quark mass, and taking the limit that $\Delta \ll m_t, m_\chi$, in the case of $\cos\theta_{\tilde t} = 0$, the decay rate of $\tilde t_1$ can be calculated analytically as
\bea\label{eq:Gammatree}
\Gamma = \frac{g_2^4 g_1^2 \Delta^8}{20160 \pi^5 m_W^4 m_t^2 m_{\tilde t_1}}. 
\eea
However, with nonzero $\cos\theta_{\tilde t}$ and $m_b \neq 0$, we need to calculate the stop decay rate numerically.

\begin{figure}[htbp]
  \centering
  \includegraphics[width=7.5cm]{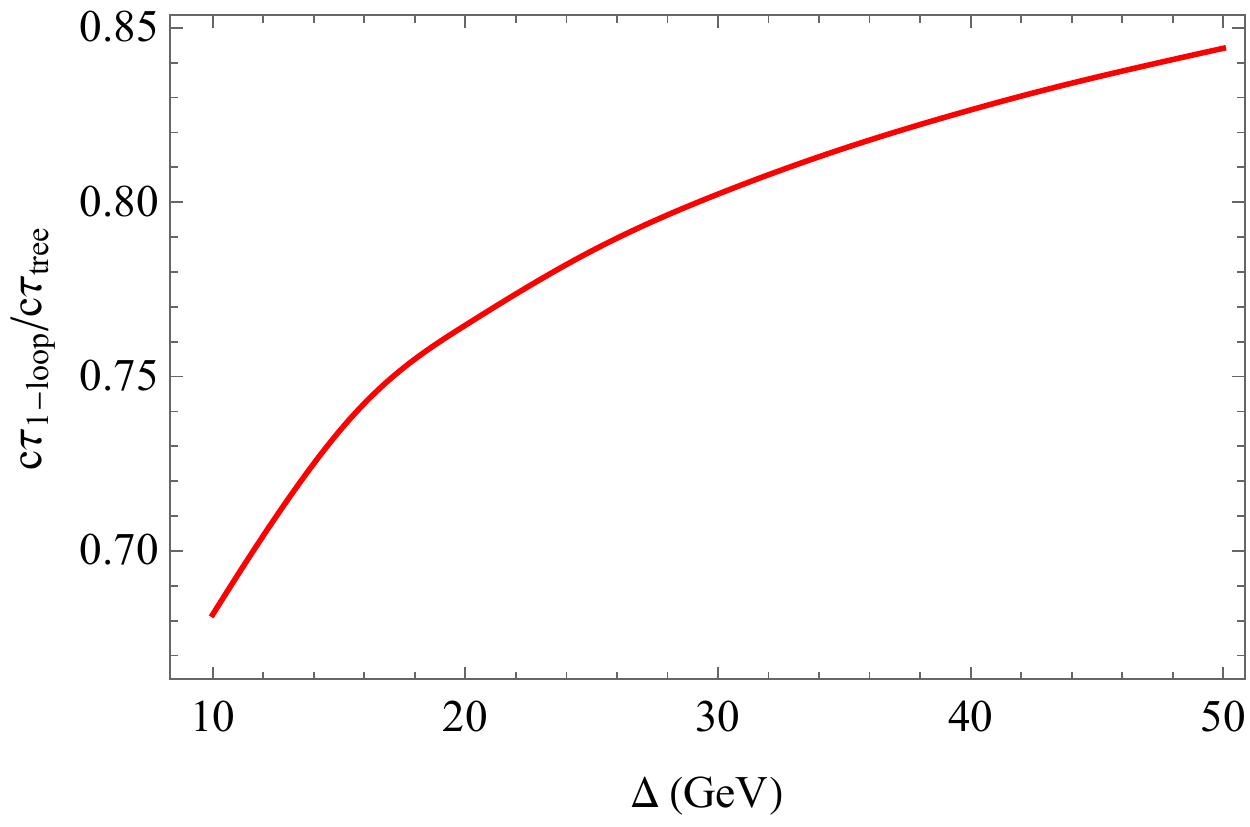}
\caption{\label{fig:ratio} The ratio of the one-loop $c\tau$ value to the leading order result as a function of $\Delta$ for $m_{\tilde t_1} = 500$ GeV. 
}
\end{figure}

In the process of the $\tilde t_1$ decay, the typical energy scale in the internal virtual particles is around $\Delta$. However, to get the effective operators, we encounter energy scales of $m_{\tilde t_1}$, $m_t$ and $m_W$, which are much larger than the small splitting $\Delta$.
Therefore, to accurately calculate the decay rate and the leptonic and hadronic branching ratios, we need to include the leading QCD corrections to the right-handed operators. The ratio of one-loop QCD corrected $c\tau$ value of purely right handed $\tilde t_1$ to the leading order value is shown in Fig.~\ref{fig:ratio}. One can see that the one-loop QCD effects induce a 20\% correction. 

To collect the leading log QCD corrections, we use the renormalization group method to crank down the energy scale from $m_{\st_1}$ to $\Delta$. 
We use the non-relativistic quantum field theory technique, similar to the heavy quark effective theory~\cite{Manohar:2000dt}, to simplify the calculation.

The QCD correction mixes the ${\cal O}_R^{(h)}$ operator with 
\bea
{\cal O}_R'^{(h)} &=& (\bar b\gamma_\mu \PL D_j)V_{ij} (\bar U_i\gamma^\mu \PL \chi  \tilde t_1) \ .
\eea
where the combination inside each bracket is a color singlet. 
Then RG evolution of their Wilson coefficients follows
\bea\label{eq:RGE1}
\frac{d}{d\log\mu} \left(\begin{array}{c}  C_R^{(h)} \\ C_R'^{(h)} \end{array}\right) = \frac{\alpha_S}{4\pi} \left(\begin{array}{cc} -5 ~&~ 3 \\ 3 ~&~ -5 \\ \end{array}\right) \left(\begin{array}{c}  C_R^{(h)} \\ C_R'^{(h)} \end{array}\right)  \ .
\eea
Whereas for the leptonic operator we have
\bea\label{eq:RGE2}
\frac{d}{d\log\mu} C_R^{(\ell)} = - \frac{\alpha_S}{\pi} C_R^{(\ell)} \ .
\eea
Eqs.~(\ref{eq:RGE1}) and (\ref{eq:RGE2}) are used to calculate the leading order RG effect from the  $m_W$ to $\Delta$. We can also include the RGE effect for the energy scale between $m_W$ and $m_t$ although this effect is not large. 
The running from $m_{\st_1}$ to $m_t$ can also be described by Eq.~(\ref{eq:RGE2}) since the color structures of the $\tilde t_1 \bar t_R \chi^0_1$ is the same as ${\cal O}_R^{(\ell)}$. 
%

In the following, we present the main steps of the derivation of the RG equations~(\ref{eq:RGE1}) and (\ref{eq:RGE2}). 
In the rest frame of the decaying $\st_1$, $\chi_1^0$ is non-relativistic. In loop diagrams the $\st_1$ internal line is also almost on-shell since $\Delta \ll m_{\st_1}$. 
%
Therefore, following Ref.~\cite{Manohar:2000dt} we redefine the $\st_1$ field as 
\bea\label{eq:redef}
\st_1 = e^{- i m_{\st_1} v\cdot x} \sT \ ,
\eea
where $v$ is the four-velocity of $\st_1$. Then the kinetic term and the mass term of $\st_1$ can be written as
\bea\label{eq:kinetic}
|\partial^\mu \st_1|^2 - m_{\st_1}^2 |\st_1|^2 = |\partial_\mu \sT|^2 + i m_{\st_1} v\cdot (\sT \partial \sT^* - \sT^* \partial \sT) \ .
\eea
Since we have already factorized the static energy of $\st_1$ by the redefinition (\ref{eq:redef}), on the righthand side of Eq.~(\ref{eq:kinetic}) the size of $\partial \sT$ is of the order of magnitude of $\Delta \sT$. Therefore, we have
\bea
- i m_{\st_1} v\cdot (\sT \partial \sT^* - \sT^* \partial \sT) \gg  |\partial_\mu \sT|^2 \ .
\eea
Therefore, we can neglect the $|\partial_\mu \sT|^2$ term. With strong interaction, the Lagrangian for $\sT$ is 
\bea\label{eq:LNR}
{\cal L_\sT} = - i m_{\st_1} v\cdot (\sT D \sT^* - \sT^* D \sT) \ ,
\eea
where $D_\mu$ is the covariant derivative for the strong interaction. With the convention that $D_\mu = \partial_\mu + i g T^a A_\mu$, where $T^a$ is the SU(3) generator, the strong interaction vertex between gluon and $\sT$ is 
\bea
- 2 i g m_{\st_1} T^a_{ij} \ .
\eea
The propagator of $\sT$ is 
\bea
\frac{i}{2m_{\st_1} v\cdot q + i\epsilon} \ ,
\eea
where $q$ is defined as $q \equiv p - m v$ and $p$ is the four momentum of $\st_1$. It can also be seen as the off-shellness of the propagator. 

In this work, we only calculate the QCD correction for the case that $\st_1$ is purely right-handed. Then the gauge-Yukawa coupling of $\sT$, $\chi_1^0$ and the top quark can be written as
\bea
y_R e^{-i m_{\st_1} v\cdot x} \sT \bar t \PL \chi_1^0 \ .
\eea 
The factor $e^{-i m_{\st_1} v\cdot x}$ ensures the energy-momentum conservation. With these Feynman rules we can calculate the loop diagrams. 

\subsection{Self energy of $\sT$}

\begin{figure}[htbp]
  \centering
  \includegraphics[width=5cm]{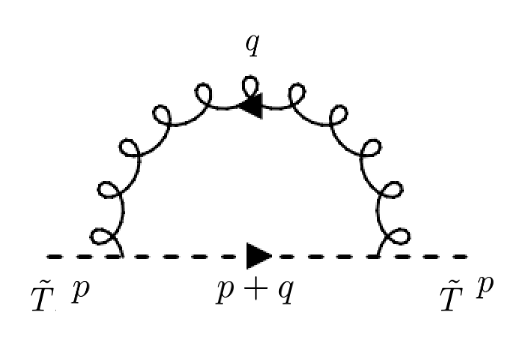}
  \includegraphics[width=5cm]{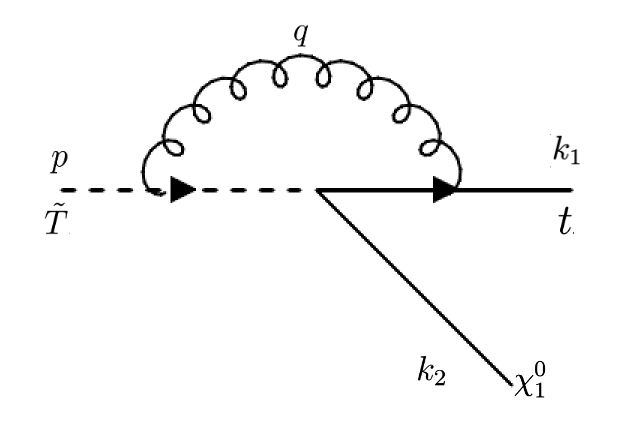}
\caption{\label{fig:self} Left: Self energy diagram of $\sT$ in non-relativistic theory. Right: One loop correction to the $\st \bar t \PL \chi^0_1$ operator. }
\end{figure}

The Feynman diagram for the $\sT$ self energy is shown in Fig.~\ref{fig:self}. We use the dimensional regularization ($d=4-2\epsilon$) to evaluate the loop diagrams. Then the self-energy of $\sT$ in the non-relativistic formalism defined above can be evaluated as 
\bea
i\Sigma(p) &=& (-2 i g m_{\st_1} v_\mu) (-2 i g m_{\st_1} v_\mu) \mu^{2\epsilon}T^a_{ik} T^a_{kj}\int \frac{d^d q}{(2\pi)^d}  \frac{i}{2 m_{\st_1}~ v\cdot(p+q)} \frac{-i}{q^2} \nn
&=& - \frac{8}{3} \delta_{ij} g^2 \mu^{2\epsilon} m_{\st_1} \int \frac{d^d q}{(2\pi)^d} \frac{1}{v\cdot (p+q)} \frac{1}{q^2 - m_{\st_1}^2} \ .
\eea
Following the steps in sec. 3.1 in \cite{Manohar:2000dt} we have
\bea
i\Sigma(p) \rightarrow -\frac{i g^2}{3 \pi^2} m_{\st_1} v \cdot p  \frac{1}{\epsilon}  + {\rm finite} \ .
\eea
Comparing to the tree-level Lagrangian in (\ref{eq:LNR}), we have at one-loop
\bea
Z_\sT = 1 + \frac{g^2}{6\pi^2 \epsilon} \ .
\eea

\subsection{Anomalous dimension of $\st \bar t \PL \chi^0_1$} 

The Feynman diagram for the QCD correction to the ${\cal O_y} = \st \bar t \PL \chi^0_1$ operator is shown in the right panel of Fig.~\ref{fig:self}. It can be evaluated as
\bea\label{eq:24}
\mu^{2\epsilon}(-i g) (-2i g m_{\st_1} v^\mu) t^a_{ik} t^a_{kj}\int \frac{d^d q}{(2\pi)^d} \bar u_t(k_1) \gamma_\mu \frac{i(\not\!k_1 + \not\!q)}{(k_1 + q)^2} \PL v_\chi(k_2) \frac{i}{2m_{\st_1} v \cdot q} \frac{-i}{q^2} \ . 
\eea
Follow the same trick in Ch. 3 of \cite{Manohar:2000dt} the divergent part of the above loop correction is
\bea\label{eq:24p}
\frac{g^2}{12\pi^2 \epsilon} \delta^{ij} \bar u_t(k_1) \PL v_\chi(k_2) \ .
\eea
Then we have 
\bea
\frac{Z_\sT^{1/2}  Z_t^{1/2}}{Z_y} - 1 = - \frac{g^2}{12 \pi^2 \epsilon} \ ,
\eea
where 
\bea
Z_t = 1 - \frac{g^2}{12 \pi^2 \epsilon} \ ,
\eea
is the renormalization factor of the relativistic quark field. Then we have
\bea
Z_y = 1 + \frac{g^2}{8\pi^2 \epsilon} \ .
\eea
The anomalous dimension of the operator is 
\bea
\gamma_y = \frac{\mu}{Z_y} \frac{d Z_y}{d\mu} = - \frac{g^2}{4 \pi^2} \ .
\eea
Then the gauge-Yukawa coupling $y_R$ can be seen as the Wilson coefficient of the operator ${\cal O}_y$ and as a result its evolution can be written as
\bea\label{eq:29}
\frac{d y_R}{d \log \mu} = - \frac{g^2}{4\pi^2} y_R \ ,
\eea
which is the same as in Eq.~(\ref{eq:RGE2}). 

\subsection{The operator $W_\mu \bar b \gamma^\mu \PL \chi_1^0 \sT$}

When the energy scale $\mu$ is below $m_t$, the top quark is integrated out and we get an effective operator ${\cal O}_W \equiv W_\mu \bar b \gamma^\mu \PL \chi_1^0 \sT$. Compared to ${\cal O}_y$, the difference is that in ${\cal O}_W$, there is a gamma matrix. However, a detailed calculation can show that this does not change the result of the one-loop anomalous dimension. The reason is that Eq.~(\ref{eq:24}) we can see that in the numerator of the top propagator, only the $\not\!q$ part contributes to the divergent part. Therefore, the coefficient of the divergent part only depends on the coefficient of  $\not\!q$. Therefore, the evolution of its Wilson coefficient $C_W$ is the same as in Eq.~(\ref{eq:29}),
\bea
\frac{d C_W}{d \log \mu} = - \frac{g^2}{4\pi^2} C_W \ .
\eea

\begin{figure}[htbp]
  \centering
  \includegraphics[width=5cm]{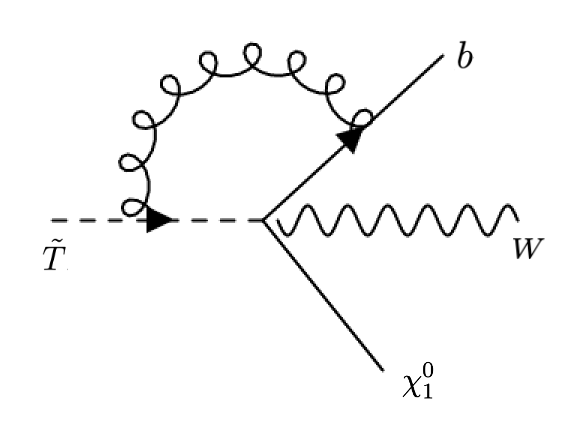}
\caption{\label{fig:OW} Left: Self energy diagram of $\sT$ in non-relativistic theory. Right: One loop correction to the $\st \bar t \PL \chi^0_1$ operator. }
\end{figure}

\subsection{Quantum corrections to ${\cal O}_R^{(\ell)}$ and ${\cal O}_R^{(h)}$}

The color structure of ${\cal O}_R^{(\ell)}$ is of no difference from ${\cal O}_W$. Therefore its evolution is the same as ${\cal O}_W$ and ${\cal O}_y$, and this is how we get Eq.~(\ref{eq:RGE2}).

\begin{figure}[htbp]
  \centering
  \includegraphics[width=14cm]{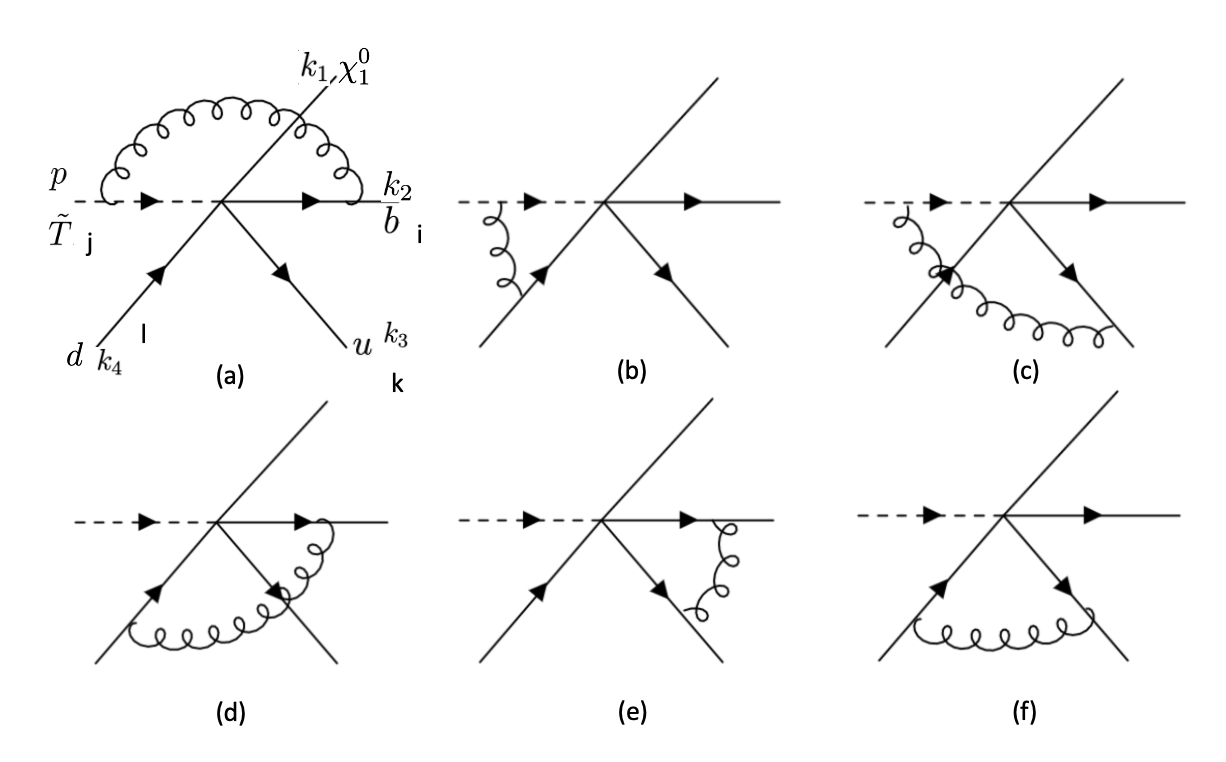}
\caption{\label{fig:OR} Left: Self energy diagram of $\sT$ in non-relativistic theory. Right: One loop correction to the $\st \bar t \PL \chi^0_1$ operator. }
\end{figure}

For ${\cal O}_R^{(h)}$ the one-loop diagrams for quantum corrections are shown in Fig.~\ref{fig:OR}. There are six diagrams contributing to the one-loop QCD correction to ${\cal O}_R$ as shown in Fig.~\ref{fig:OR}. One can see that the result of diagram (a) is the same as the one-loop result in Fig.~\ref{fig:OW}, and therefore its divergent part has the same coefficient as in (\ref{eq:24p}), and can be written as
\bea
\frac{g^2}{12\pi^2 \epsilon} \delta^{ij} \delta^{kl} \bar u_b \gamma^\mu \PL \chi \sT \bar u_u \gamma_\mu \PL v_d \ .
\eea
Explicit calculation of diagrams (b) and (c) show that their divergent parts are canceled with each other. The loops in diagrams (d) and (e) are only composed of light quarks and therefore can be done with standard calculation. In these two diagrams the exchange of the gluon mixes the color indices. The results are, for diagram (d)
\bea\label{eq:32}
\frac{g^2}{16\pi^2 \epsilon} t^a_{ij} t^a_{kl} \bar u_b \gamma^\mu \PL \chi \sT \bar u_u \gamma_\mu \PL v_d \ ,
\eea
and for diagram (e)
\bea\label{eq:33}
-\frac{4g^2}{16\pi^2 \epsilon} t^a_{ij} t^a_{kl} \bar u_b \gamma^\mu \PL \chi \sT \bar u_u \gamma_\mu \PL v_d \ .
\eea
With the formula $t^a_{ij} t^a_{kl} = \frac{1}{2} \delta_{il}\delta_{jk} - \frac{1}{6} \delta_{ij} \delta_{kl}$, we can see that the operator ${\cal O}'^{(h)}_R$ appears from the quantum correction. 
For diagram (f), from the well-known result of Ward identity, we know that its contribution to the correction of ${\cal O}^{(h)}_R$ is canceled by the contributions from the wave function renormalization. Collecting the above results, we get the renormalization factors for $O_R^{(h)}$ and $O_R'^{(h)}$ 
\bea
Z_R^{(h)} = \left(\begin{array}{cc}  1 + \frac{5g^2}{32\pi^2\epsilon} ~&~  - \frac{3g^2}{32\pi^2 \epsilon}   \\ - \frac{3g^2}{32\pi^2 \epsilon} ~&~   1 + \frac{5g^2}{32\pi^2\epsilon} \\ \end{array}\right) \ . 
\eea
Then we can get the anomalous dimensions following the standard procedure that
\bea
\gamma_R^{(h)} = \left(\begin{array}{cc}  \frac{5g^2}{16\pi^2} ~&~  - \frac{3g^2}{16\pi^2 }   \\ - \frac{3g^2}{16\pi^2 } ~&~  \frac{5g^2}{16\pi^2 } \\ \end{array}\right) \ .
\eea
This leads to the evolution of the Wilson coefficients in Eq.~(\ref{eq:RGE1}).

\section{Details of Open Data Treatment}
\label{app:cmsdetails}

\subsection{Selection efficiency of track impact parameter significance}
\label{sec:tkeff}




Each displaced track needs to satisfy an impact parameter significance selection $|d_{xy}|/\sigma_{d_{xy}}>4$.
To estimate the efficiency of this selection, we need to know the $\sigma_{d_{xy}}$ distribution, which is a pure detector effect and cannot be obtained without full detector simulation. 
For a simple estimate, we use the $\sigma_{d_{xy}}$ distribution of the high purity prompt tracks, selected with $|d_{xy}|<250~\rm\mu m$ and $|\eta|<2.4$ of the CMS Open Data, to approximate the true distribution. 
We use prompt tracks because they are mostly truly identified, similar to our tracks from top squark and B meson decays. 
As the error of a distance between two points is dependent only on that of the two endpoints, we expect the $\sigma_{d_{xy}}$ distribution to have insignificant $d_{xy}$ dependence for physically originated tracks.  

\begin{figure}[htbp]	
  \centering
  \includegraphics[width=9.5cm]{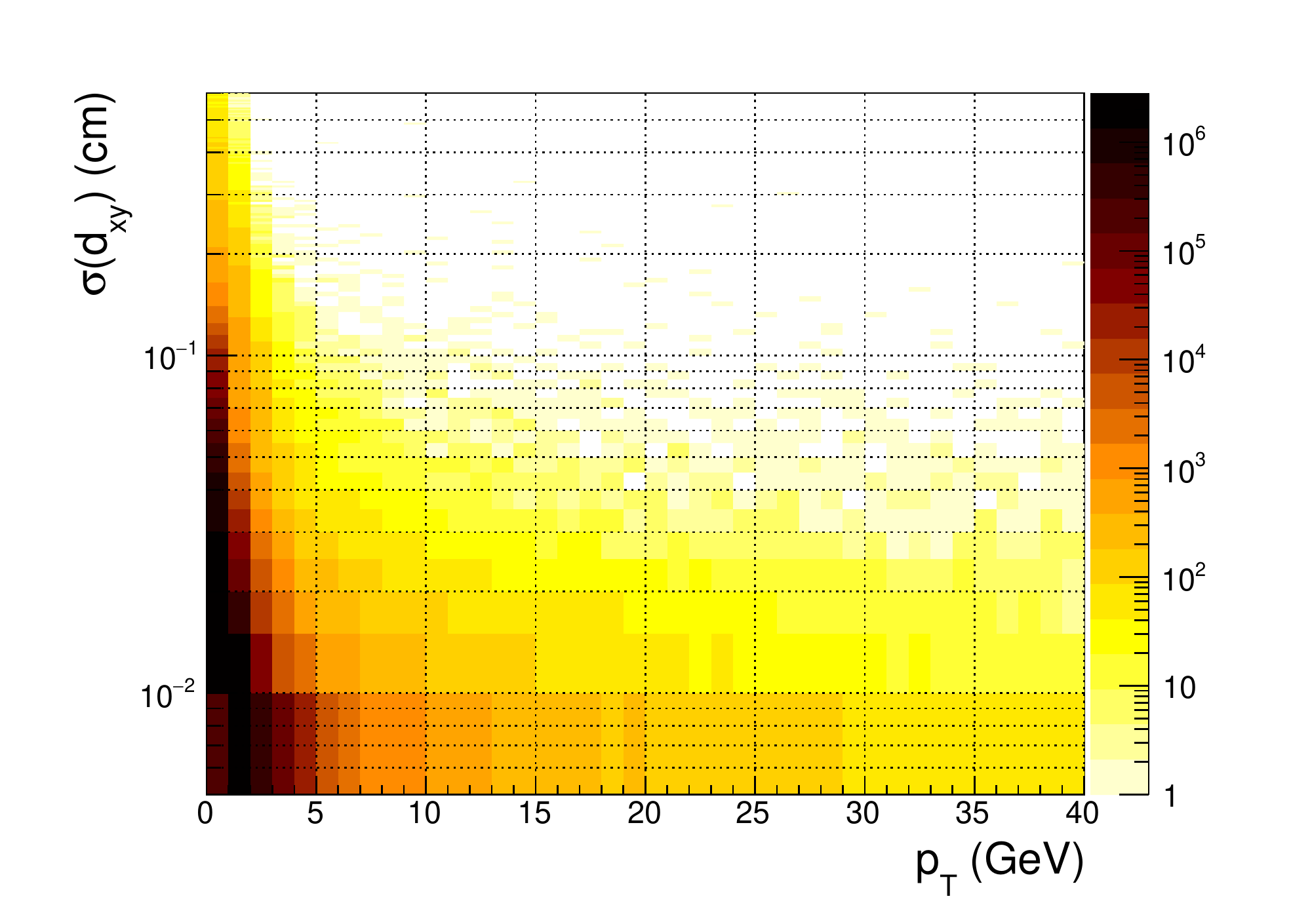}
\caption{\label{appfig:tkeff} Two dimensional $\pt-\sigma_{d_{xy}}$ histogram used for the computation of the track impact parameter significance selection efficiency. }
\end{figure}

Based on the $\pt-\sigma_{d_{xy}}$ histogram of the selected high purity prompt tracks, we calculate the selection efficiency in $\pt$ bins from $0-40$ GeV, with a bin-width of 1 GeV. 
For a track with $\pt^0$ and $d_{xy}^0$, the efficiency is computed as the ratio of the number of tracks satisfying $\sigma_{d_{xy}}<|d_{xy}^0|/4$ and the total number of tracks in that $\pt$ bin, i.e., $N(\sigma_{d_{xy}}<|d_{xy}^0|/4~|~\pt^0)/N(\sigma_{d_{xy}}>0~|~\pt^0)$.
We show in Fig.~\ref{appfig:tkeff} the two dimensional $\pt-\sigma_{d_{xy}}$ histogram used for this computation. 

\subsection{Displaced vertex efficiency}
\label{app:vtx}

Displaced vertices in simulated events should be reconstructed using the same algorithm as for data events. Without the capability of doing full detector simulation for the signal samples, we extract the vertex efficiency using a fully simulated 8 TeV $t\bar{t}$ sample obtained from the CMS Open Data portal.
Our method is viable because both generator- and detector-level information are available in the fully simulated AOD sample. 
At the generator level, we select hadronically decaying $B_0$ or $\bar{B}_0$ mesons whose vertex position is at least $0.5$ mm, but smaller than $18$ mm 
from the beam-line, and have energies within 10 to 30 GeV. Here, we make a harder upper cut on the signal displacement to be on the conservative side. 
We further categorize these candidates by their number of MC charged tracks $N_{gen,tk}$, where the tracks satisfy $d_{xy}>0.5$ mm, $\pt>1$ GeV and $|\eta|<2.4$ and are not coming from tertiary vertices.
In this way, we prepared several categories of signal-like displaced vertices. 

To calculate the vertex efficiency in each category, we reconstruct displaced vertices in the same way as we have done for data and check if a signal-like vertex candidate is reconstructed as a displaced one. 
For the latter purpose, one needs to introduce some matching criteria. 
We consider two methods to ensure the reliability of our computation. 
In the first method, hereafter track fraction (TF) method, we consider matching criteria based on the fraction of tracks associated with the reconstructed displaced vertex from B-meson decays. 
First, we consider the following criteria for matching a track:
\begin{itemize}
\item $\Delta R(p_1,p_2)<0.1$, where $\Delta R(p_1,p_2)\equiv\sqrt{(\eta_1-\eta_2)^2+(\phi_1-\phi_2)^2}$ is the $\eta-\phi$ separation between the candidates; 
\item absolute value of the $\pt$ difference between the two tracks is smaller than 10\%;
\item distance in the transverse plane is smaller than $0.5$ mm.
\end{itemize}
We have also checked that the matching quality is not sensitive to these parameters.
Based on the matched tracks, we consider the following criteria for matching a displaced vertex:
\begin{itemize}
\item the fraction of matched tracks among all iterated reconstructed tracks is higher than 0.4;
\item the reconstructed vertex position is at least $0.1$ mm but smaller than $20$ mm from the beam-line;
\item invariant mass for the sum of iterated reconstructed tracks is smaller than 6.5.
\end{itemize}
The requirements of the number of iterated tracks and the vertex position to be at least $0.1$ mm from the beam-line eliminate misassociations to primary vertices.

The second method, hereafter the vertex distance (VD) method, we do not match tracks but simply require that the distance between simulated vertex and the reconstructed vertex to be smaller than its one sigma uncertainty, which is computed as:
\begin{eqnarray}
\sigma(|\vect{d}_{VV}|) = \dfrac{1}{|\vect{d}_{VV}|} \sqrt{\sum_{i,j} (\vect{d}_{VV})_i c_{ij} (\vect{d}_{VV})_j }. 
\end{eqnarray}
where $\vect{d}_{VV}$ is the vector from the position of the simulated vertex to that of the reconstructed one, and $c_{ij}$ is the
covariant matrix of the vertex error where $i,j$ runes over $x,y,z$.

We summarize results from the two methods in Table~\ref{tab:vtxeff}. 
We adopt results using the TF method for the analysis in this study and consider the VD method proof of concept. 
We find the matched candidates have $\gtrsim 60$\% overlap for all the catalogs.
In Fig.~\ref{fig:vtxff}, we check property of the signal-like categories by comparing their $H^V_T$ distribution to that of a signal benchmark with $\Delta=20~$GeV and $m_{\tilde{t}_1}=360~$GeV, in the one-vertex control region.

As shown in Table~\ref{tab:vtxeff}, the vertex efficiency is not monotonically increasing with $N_{gen,tk}$. The reason for this is due to contamination of additional tracks from the collider environment. 
To demonstrate that, we put extra matching conditions $\Delta R<0.1$, $|\Delta\pt/\pt|<0.05$ on the set of reconstructed tracks and reconstruct displaced vertices based on these purified tracks.
As expected, the efficiency increased significantly, reaching 85.1\% in the TF method for the $N_{gen,tk}=4$ catalog.

\begin{table}[bthp]
\begin{center}
\begin{tabular}{c|c|c|c|c|c}
\hline
\hline
Catalog                                  & $N_{gen,tk}=2$ & $N_{gen,tk}=3$ & $N_{gen,tk}=4$ & $N_{gen,tk}=5$ & $N_{gen,tk}\geq 6$  \\
\hline
Efficiency from TF (\%)                  &   $23.8\pm 0.4$&   $36.6\pm 1.0$&  $46.1\pm 2.9$ &  $45.3\pm 6.2$ &  $32.4\pm 10.8$  \\
Efficiency from VD (\%)                  &   $17.5\pm 0.3$&   $25.7\pm 1.0$&  $32.6\pm 2.4$ &  $32.6\pm 5.0$ &  $40.5\pm 12.4$  \\
Overlapping fraction (\%)                &   59.7         &   62.0         &    64.3        &    70.5        &    83.3  \\
Vertex error ($\rm \mu m$)               &   173          &   170          &    164         &    175         &    155   \\
Vertex error RMS ($\rm \mu m$)           &   110          &   110          &    103         &    119         &    94.5  \\
\hline
Probability of passing $N_{vtx,tk}\geq$2 &   1.0          &   1.0          &    1.0         &    1.0         &    1.0   \\
Probability of passing $N_{vtx,tk}\geq$3 &   0.61         &   0.78         &    0.83        &    0.82        &    0.83  \\
Probability of passing $N_{vtx,tk}\geq$4 &   0.23         &   0.39         &    0.54        &    0.64        &    0.58  \\
\hline
\hline
\end{tabular}
\caption{\label{tab:vtxeff} Efficiencies and errors of the reconstructed displaced vertices. The overlapping fraction is computed as the ratio of the number of vertices matched in both methods to that matched only in the TF method. }
\end{center}
\end{table}

\subsection{Merging MC vertices}
\label{app:vtxcomb}

In the experiment, two displaced vertices can not be resolved if they are too close. 
To take into account this effect, we merge nearby simulated vertices in each event before applying the efficiencies in Table~\ref{tab:vtxeff}.
We merge nearby displaced vertices through the following steps: 
\begin{itemize}
\item For all simulated tracks associated with displaced vertices, apply the track efficiencies. Count the number of selected tracks associated with each vertex.  
\item Assign errors to these vertices. Based on the number of selected tracks, the error is generated from a Gaussian random distribution considering error and RMS values in Table~\ref{tab:vtxeff}. 
As we have chosen the impact parameter selection efficiency 
to mimic that of the impact parameter significance
selection one, we expect $N_{{\sc pythia},tk}\approx N_{gen,tk}$ and do not distinguish them.
\item Merge nearby vertices into one if their distance is smaller than the sum of their errors in quadrature. 
The new position is calculated as an average of candidate vertices weighted by their number of selected tracks. 
The error of the merged vertex is taken as those of the candidates summed in quadrature. 
\item Based on the number of $N_{{\sc pythia},tk}$ of the merged vertices, apply the vertex identification and reconstruction efficiency in Table~\ref{tab:vtxeff}.
\end{itemize}

\clearpage

\subsection{Supplementary figures}

\begin{figure}[htbp]
  \centering
  \includegraphics[width=15cm]{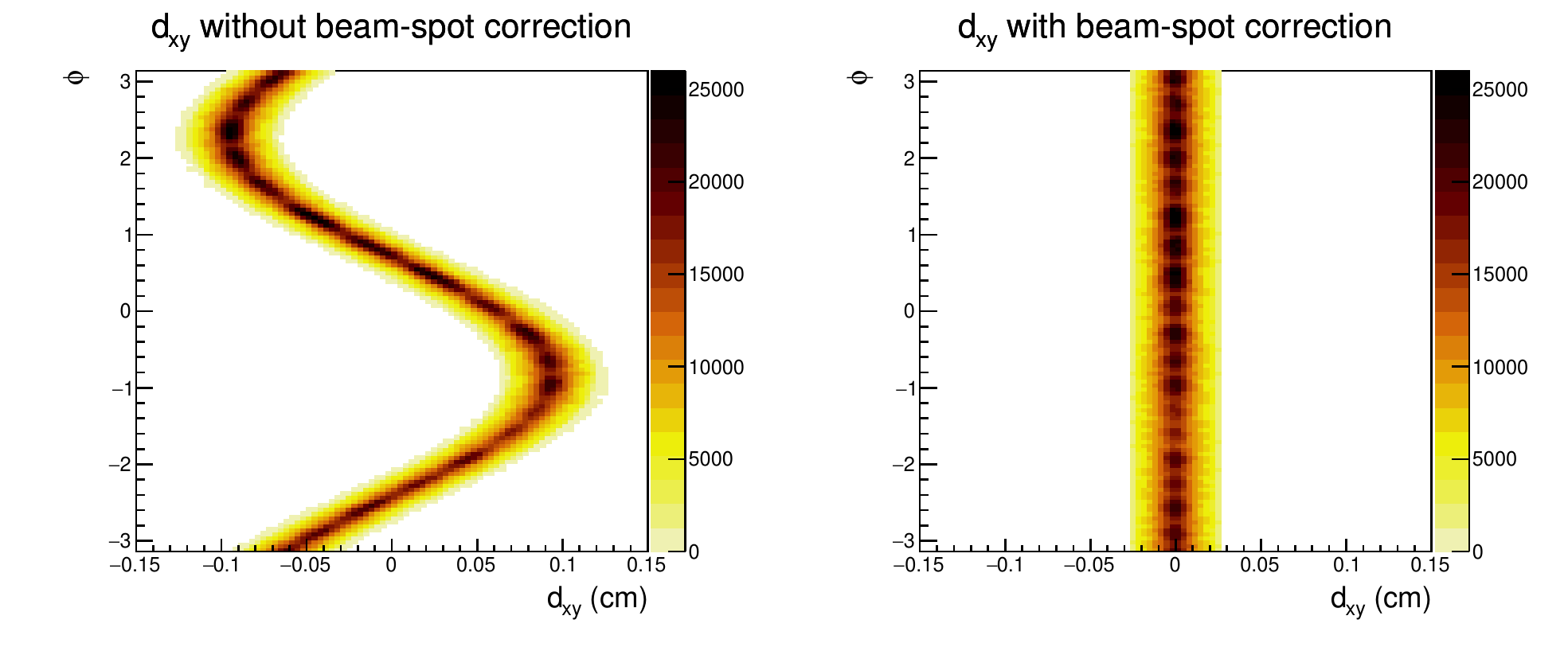}
\caption{\label{fig:bscorr} Transverse track impact parameter distribution, computed with (right) and without (left) beam-spot correction. The figure is made using tracks from 10k data events.} 
\end{figure}

In Fig.~\ref{fig:bscorr}, we show the importance of beam-spot correction.
Performing this correction, we found the distribution to be uniform in the $\phi$ direction.

\clearpage

\begin{figure}[htbp]
  \centering
  \includegraphics[width=7.5cm]{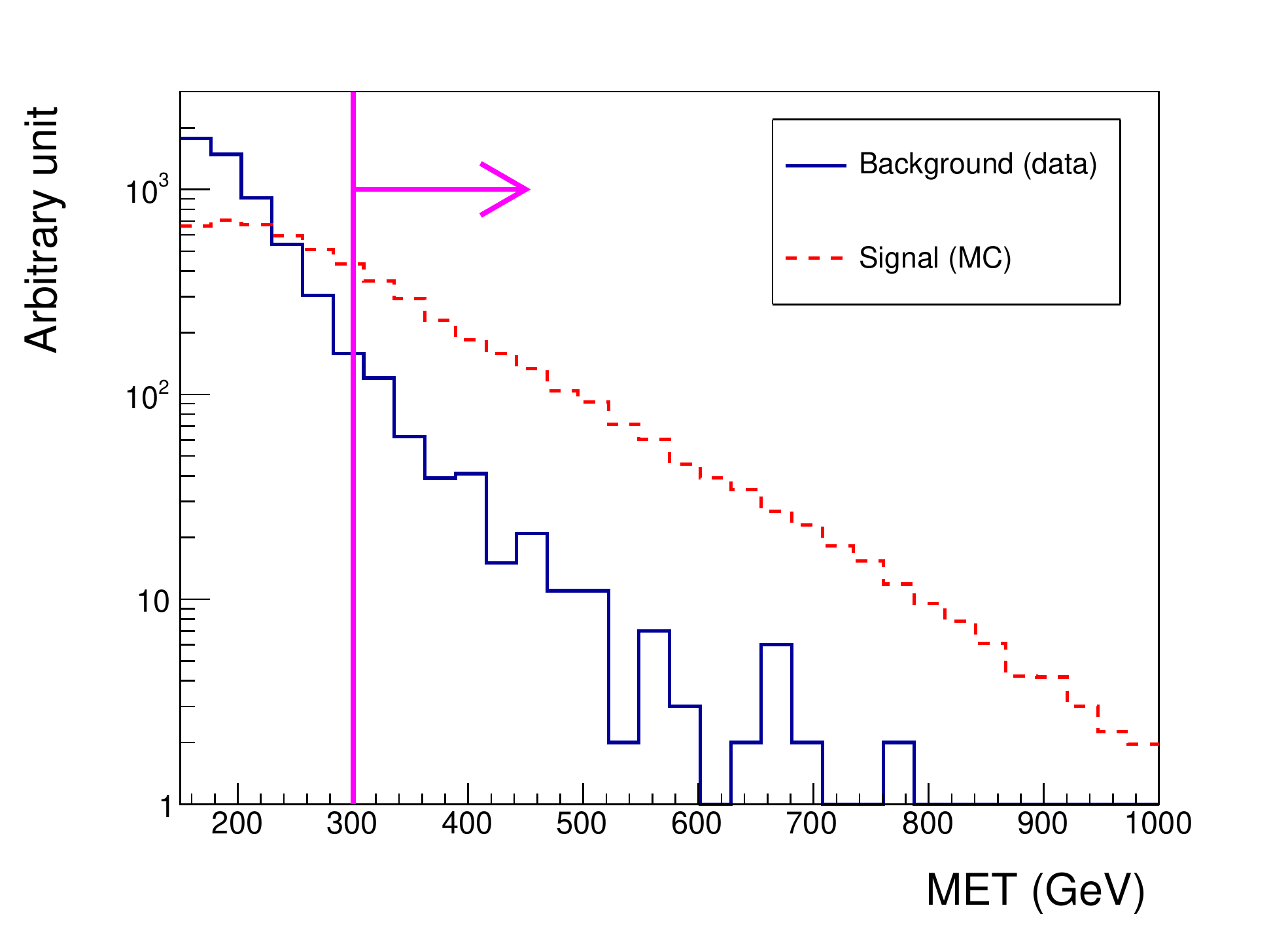}
  \includegraphics[width=7.5cm]{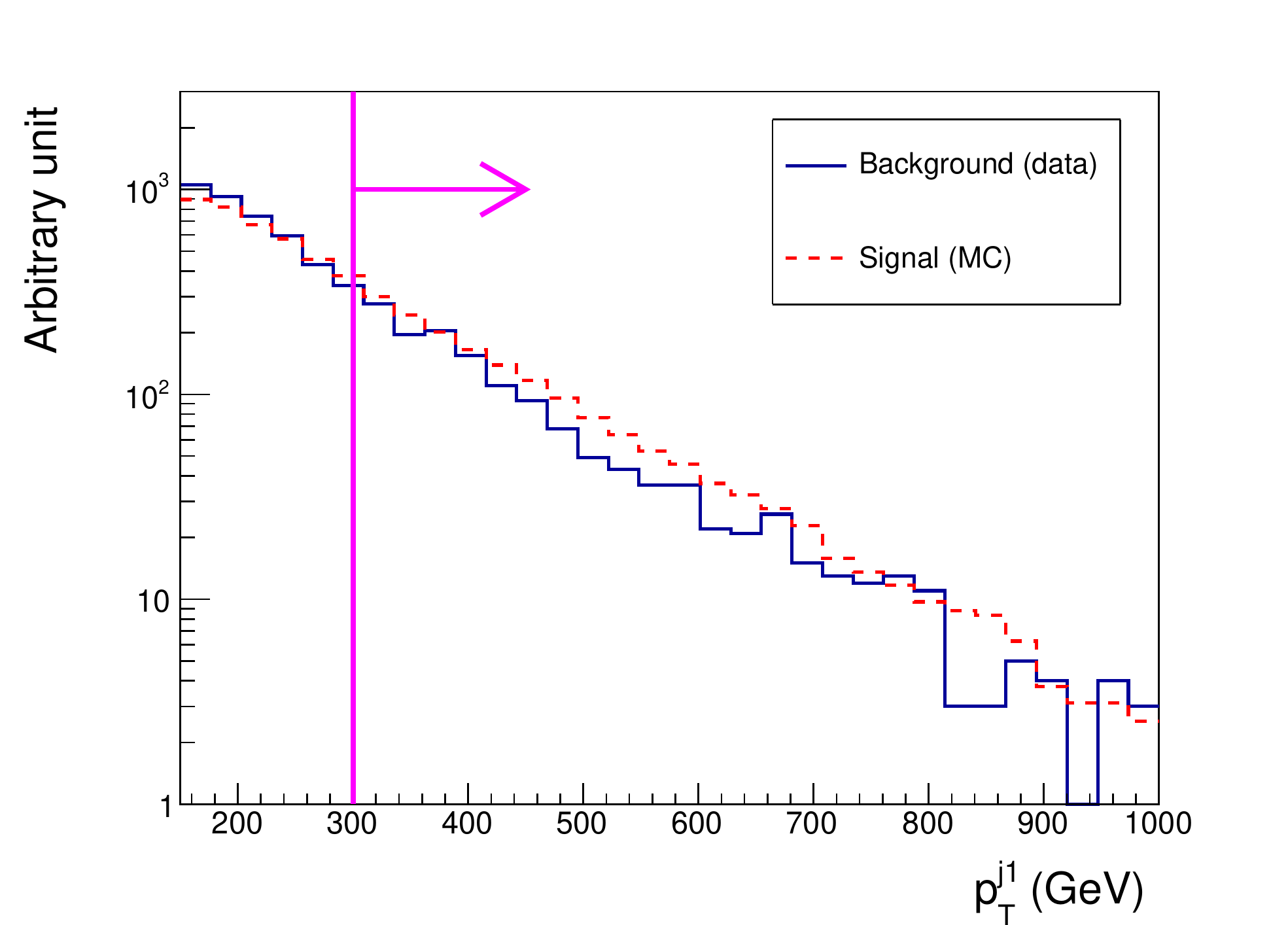} \\
  \includegraphics[width=7.5cm]{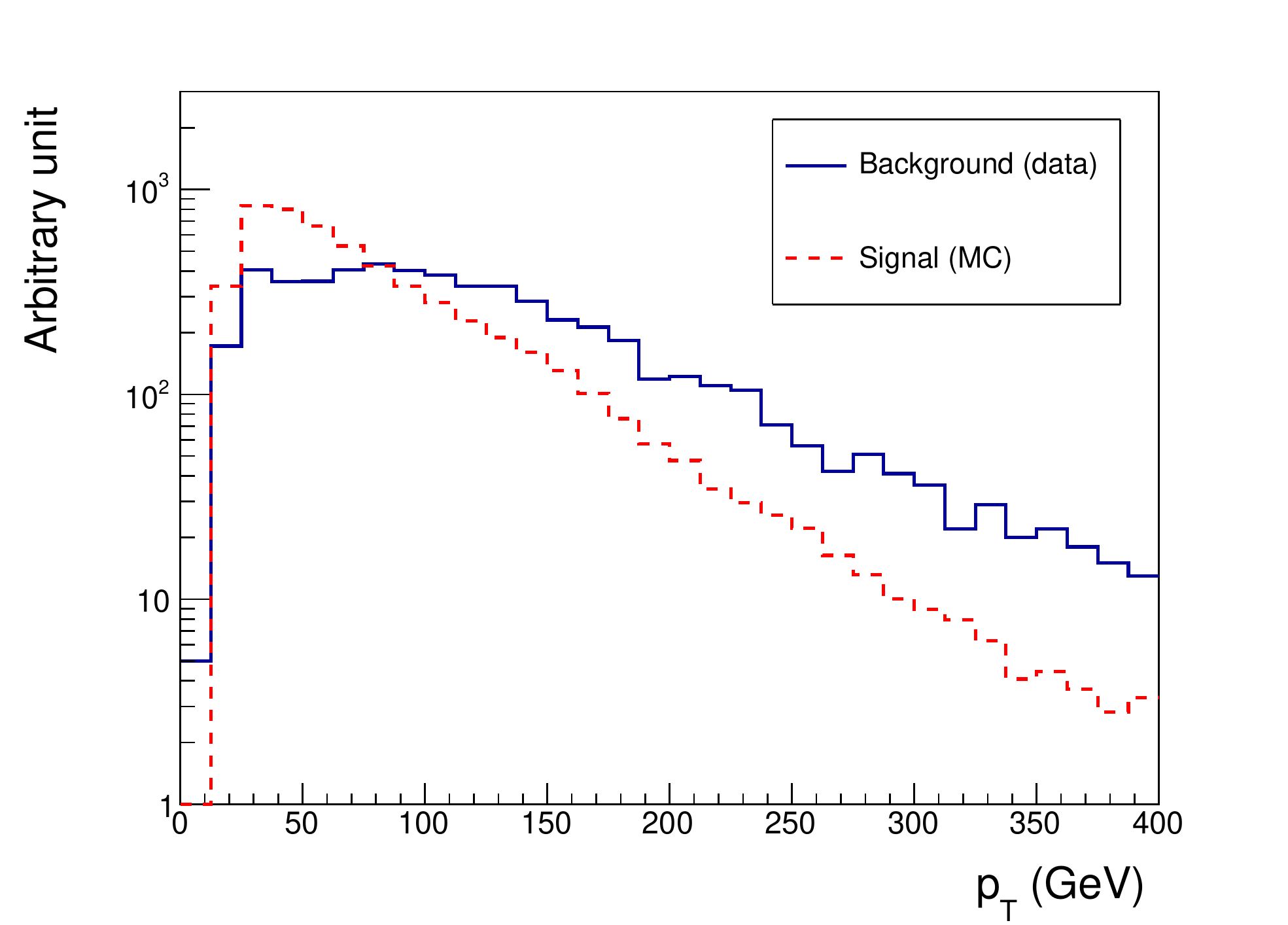}
  \includegraphics[width=7.5cm]{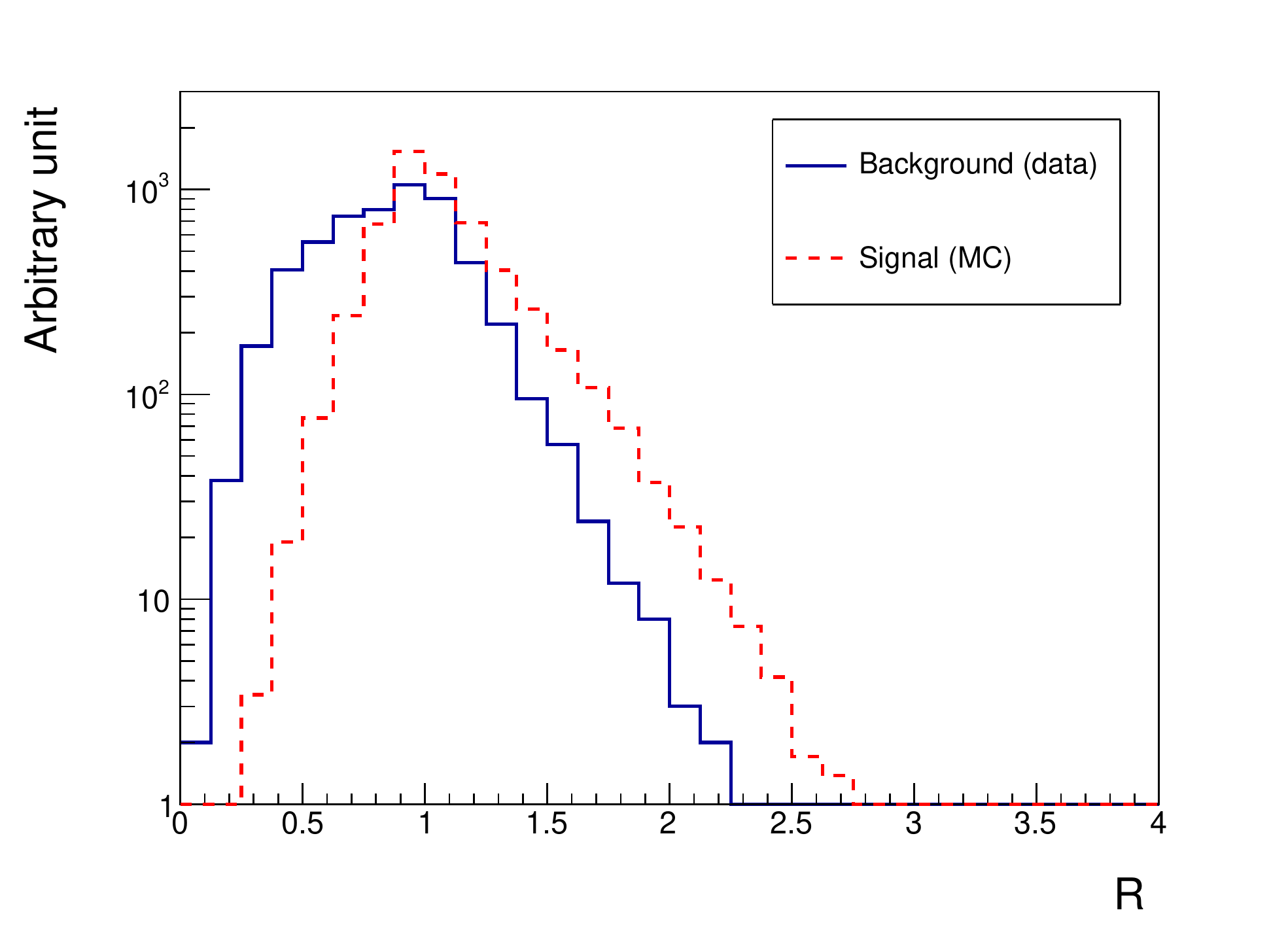} \\
  \includegraphics[width=7.5cm]{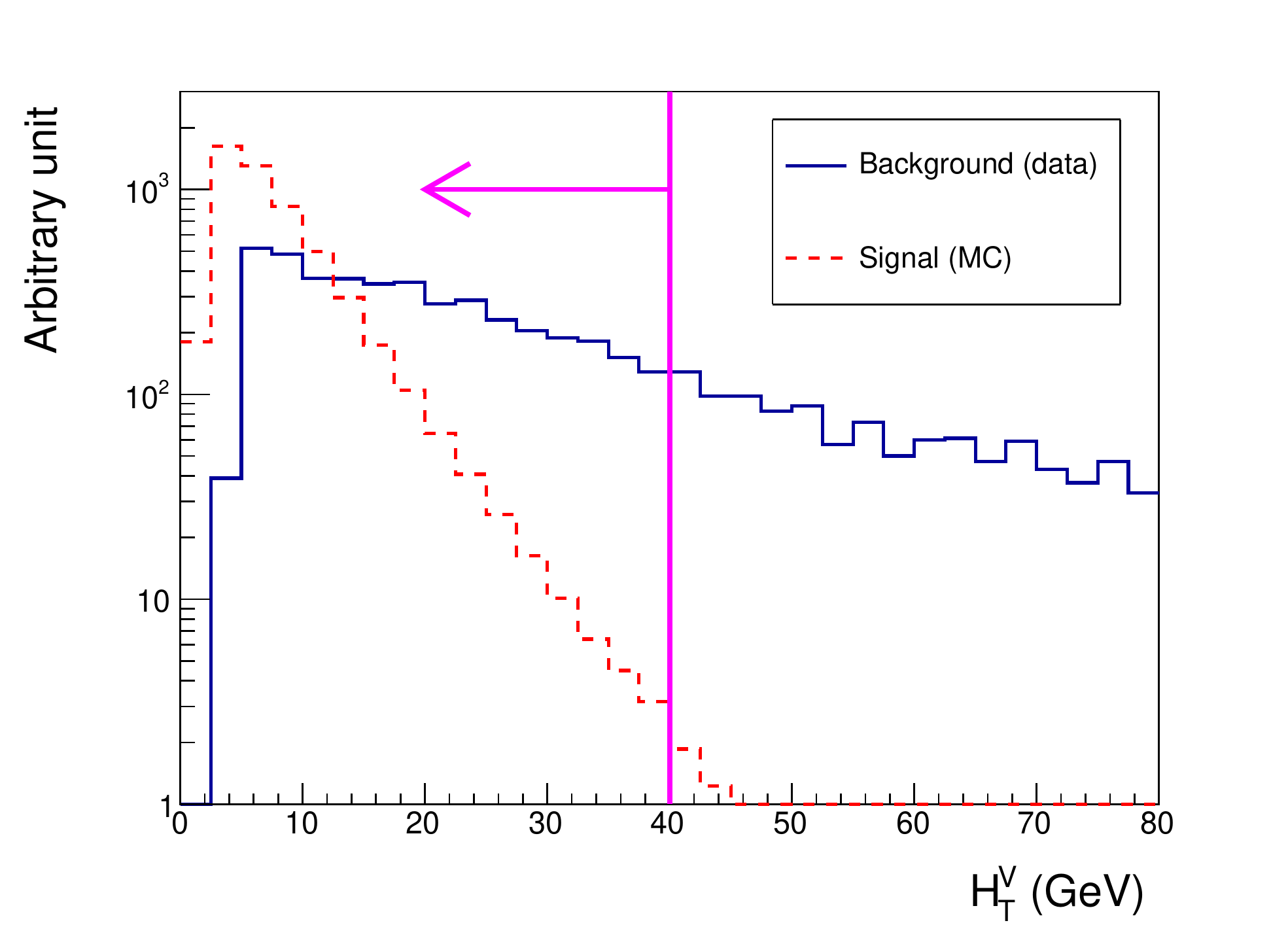}
  \includegraphics[width=7.5cm]{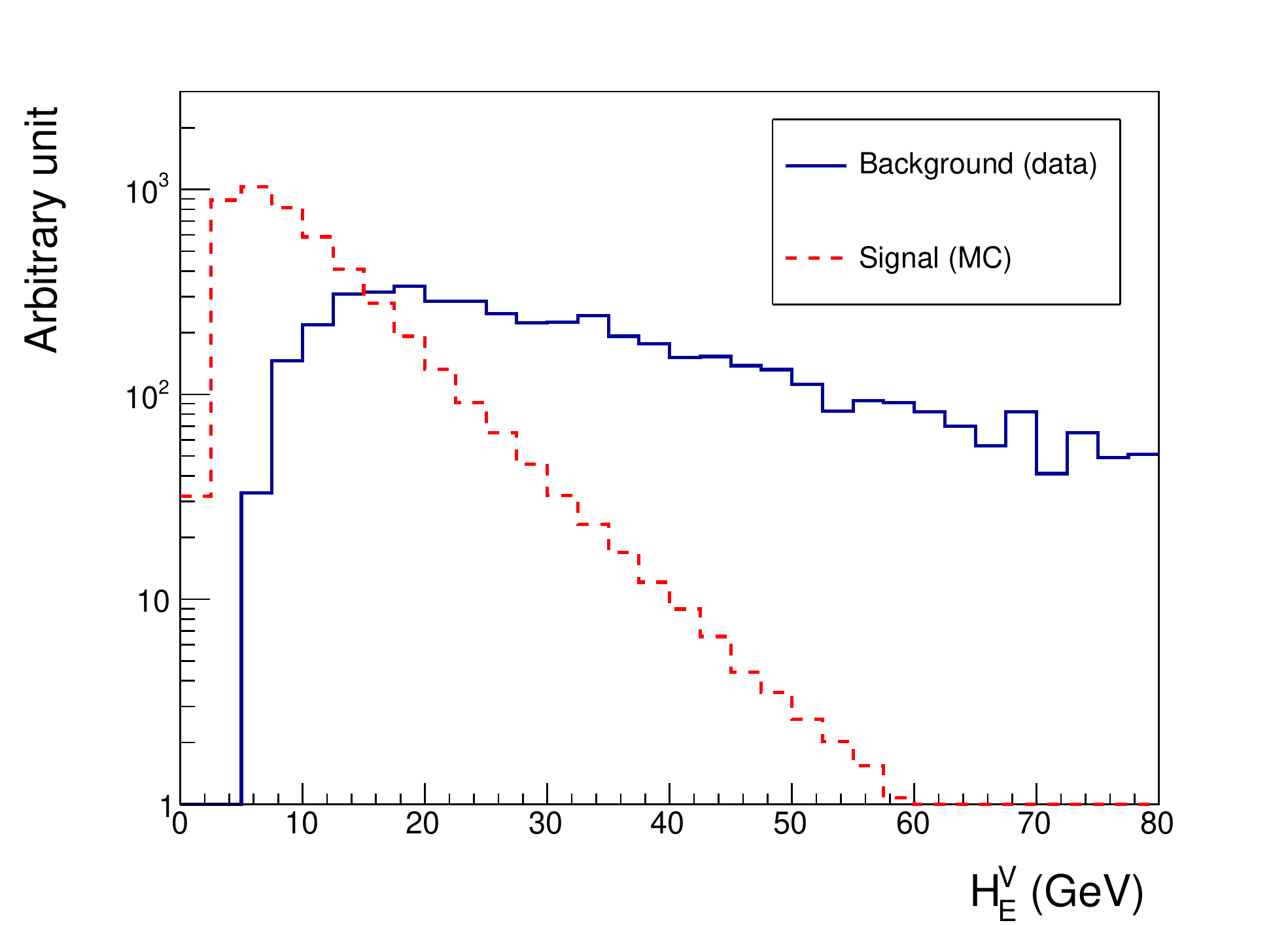}
\caption{\label{fig:ctlplots} 
Comparisons of signal and background kinematic features. 
The signal sample is generated considering a top squark mass of 360 GeV and a mass gap of 20 GeV.
The background sample is taken from data in the one-vertex control region, where we expect the signal contribution to be negligible. 
Their differences reveal features that may be exploited to enhance the signal significance. 
In this analysis, we cannot exploit all features due to limited statistics. 
}
\end{figure}

In Fig.\ref{fig:ctlplots}, we compare the signal and data features in the one-vertex control region.  
The one-vertex control region is defined by requiring one displaced vertex with $N_{vtx,tk}\geq$4, MET larger than 150 GeV, and transverse momentum of the ISR jet higher than 150 GeV.

\clearpage

\section{Current limits with 13 TeV Data}

\begin{figure*}[tb]
  \centering
  \includegraphics[width=8cm]{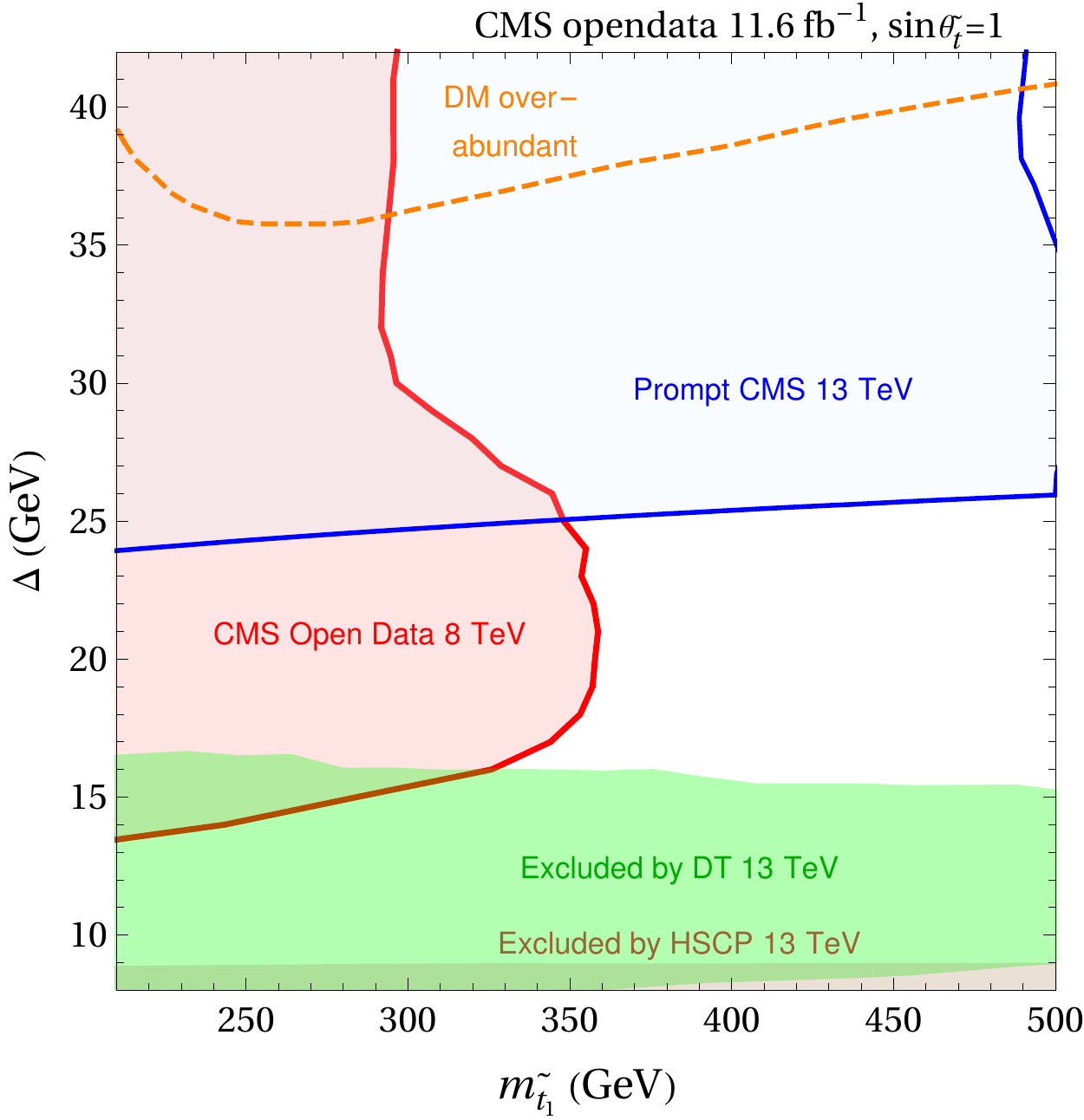}
  \includegraphics[width=8cm]{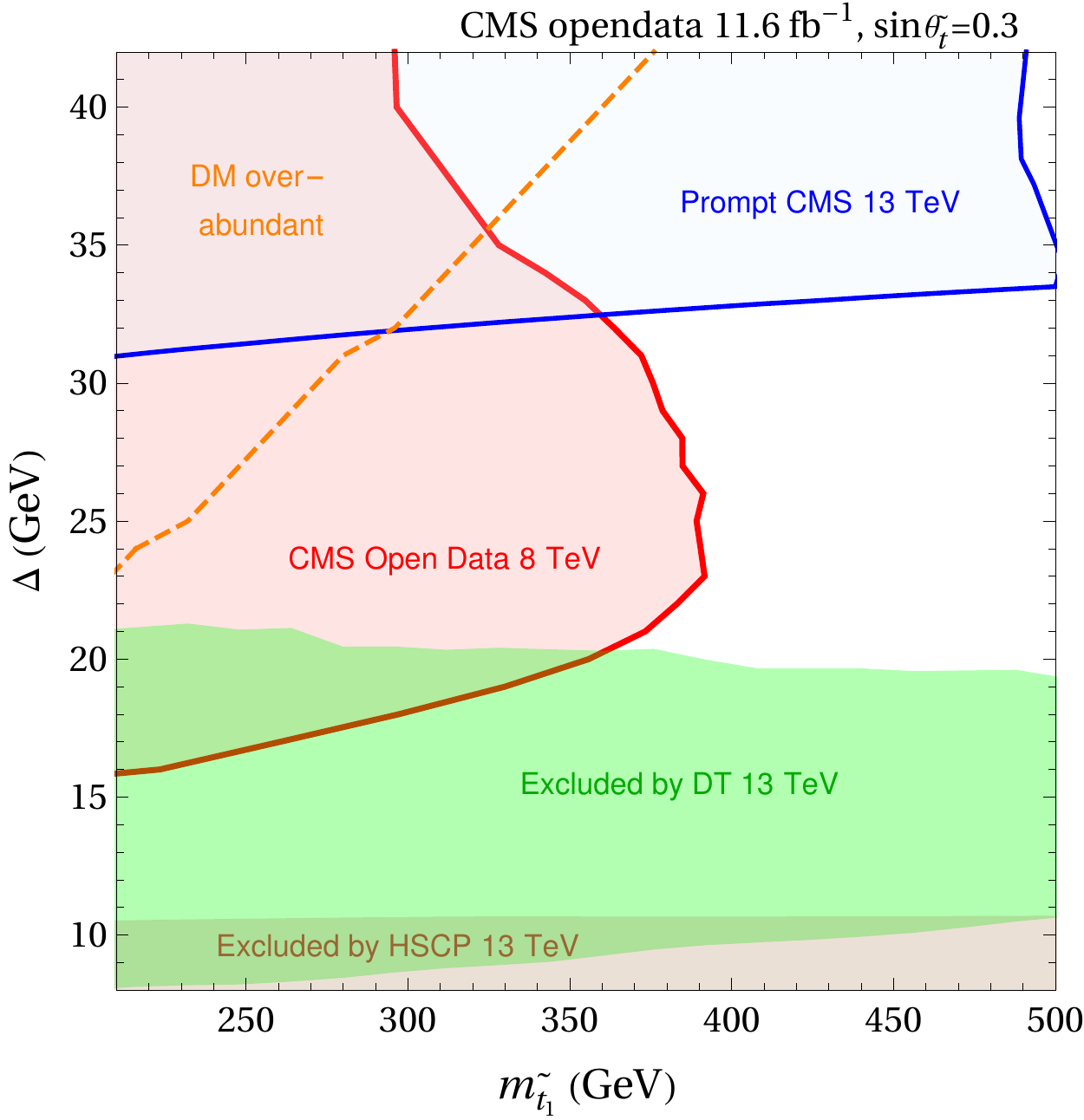}
\caption{\label{fig:limits13} Same as Fig.\ref{fig:limits}, but with existing exclusion limits replaced by the 13 TeV ones. 
The 13 TeV CMS limits come from Ref.~\cite{Sirunyan:2018omt}($\tilde{t}\to b f f' \tilde{\chi}_1^0$ channel, MVA approach). 
The limits from disappearing track (DT)~\cite{Sirunyan:2018ldc} and heavy stable charged particle (HSCP)~\cite{Khachatryan:2016sfv,Aaboud:2019trc} searches are also replaced by the ones re-interpreted from 13 TeV analyses, using the framework in Ref.~\cite{Liu:2015bma,Kraml:2013mwa,Ambrogi:2018ujg}. 
The orange-dashed curves show the contours above which the models would lead to over-abundant DM. The corresponding relic densities, for the $\sin\theta_{\tilde{t}}=1$ case, is taken from Ref.~\cite{Keung:2017kot} with the bound state effects included; for the $\sin\theta_{\tilde{t}}=0.3$ case, is computed using the {\sc MadDM} program~\cite{Backovic:2013dpa,Arina:2020kko}, considering a sbottom mass of 1 TeV. 
}
\end{figure*}

In the text, for the purpose of a fair comparison, we show the obtained limit contour with the 8 TeV ones. 
To understand the current status of the constraints, we show in Fig.~\ref{fig:limits13} a comparison of our limit with the 13 TeV ones.  
The 13 TeV CMS limit is taken from Ref.~\cite{Sirunyan:2018omt}($\tilde{t}\to b f f' \tilde{\chi}_1^0$ channel, MVA approach) which exploited a data set corresponding an integrated luminosity of 35.9 $fb^{-1}$.  
As in Fig.~\ref{fig:limits}, we also include limits from DT and HSCP experiments. 
These limits are recasted from 13 TeV experiments~\cite{Sirunyan:2018ldc,Khachatryan:2016sfv,Aaboud:2019trc}, using the framework in Ref.~\cite{Liu:2015bma,Kraml:2013mwa,Ambrogi:2018ujg}. 
From Fig.~\ref{fig:limits13}, we see the 13 TeV experiments exclude the majority of parameter space that corresponds to the prompt or the stable heavy charged particle. 
Similar to our reinterpretation of the prompt search at 8 TeV, we choose 0.2~mm as the cut for prompt searches. Note that the lifetime changes as the eighth power of the mass splitting; hence the prompt search lose sensitivity rapidly as we decrease the splitting. Admittedly, if the exact prompt search parameters are known, one can perform a more faithful reinterpretation of the 13 TeV prompt analysis. In this case, we would anticipate a smooth transition, rather than the abrupt kink in the sensitivity curve near our $c\tau$ cut in the current figure, the prompt search probably will start loosing efficiency at $c\tau$ of 0.5~mm, as in the case of Higgs exotic decay experimental reinterpretation~\cite{ATLAS:2018pvw}. The prompt analysis loses signal efficiency by more than one order of magnitude from 0.5~mm to 2~mm.
Our analysis approach provides a promising test of the parameter region in the middle, which corresponds to the existence of the long-lived decaying particle. 

\end{document}